\documentstyle[preprint,tighten,aps,floats,epsf,psfig,amssymb]{revtex}

\begin{document}

\def\spose#1{\hbox to 0pt{#1\hss}}
\def\ltapprox{\mathrel{\spose{\lower 3pt\hbox{$\mathchar"218$}}
 \raise 2.0pt\hbox{$\mathchar"13C$}}}
\def\gtapprox{\mathrel{\spose{\lower 3pt\hbox{$\mathchar"218$}}
 \raise 2.0pt\hbox{$\mathchar"13E$}}}

\draft

\title{ 
Critical structure factors of bilinear fields in O($N$)-vector models. 
}
\author{
  \\
  {\small Pasquale Calabrese,${}^{a,}$\cite{PC-email} 
          Andrea Pelissetto,${}^{b,}$\cite{AP-email}  
          Ettore Vicari${}^{c,}$\cite{EV-email}  
}
\\
  {\small\it ${}^a$ Scuola Normale Superiore and
              INFN, Sezione di Pisa,
              I-56100 Pisa, ITALIA} \\
  {\small\it ${}^b$ Dipartimento di Fisica, Universit\`a di Roma La Sapienza} 
         \\
  {\small\it and 
              INFN, Sezione di Roma,
              I-00185 Roma, ITALIA} \\
  {\small\it ${}^c$ Dipartimento di Fisica, Universit\`a di Pisa and 
              INFN, Sezione di Pisa,
              I-56100 Pisa, ITALIA} \\
}

\maketitle

\begin{abstract}
We compute
the two-point correlation functions of general quadratic operators
in the high-temperature phase of the 
 three-dimensional $O(N)$ vector model by using 
field-theoretical methods.
In particular, we study the small- and large-momentum behavior
of the corresponding scaling functions, 
and give general interpolation formulae based on a dispersive approach.
Moreover, we determine the crossover exponent $\phi_T$ associated with the 
traceless tensorial quadratic field, 
by computing and analyzing its six-loop 
perturbative expansion in fixed dimension. 
We find: $\phi_T=1.184(12)$, $\phi_T=1.271(21)$, and 
$\phi_T=1.40(4)$ for $N=2,3,5$ respectively.
\end{abstract}



\section{Introduction}

In nature many physical systems undergo phase transitions belonging
to universality classes of the  O($N$) vector models. 
Their universal critical properties can be determined theoretically 
by considering the $\phi^4$ Hamiltonian
\begin{equation}
{\cal H}= \int d^dx\left[ {1\over 2}\partial_\mu \vec{\phi}\cdot \partial_\mu \vec{\phi}
+ {1\over 2} r \vec{\phi}\cdot \vec{\phi} + {1\over 4!}u
(\vec{\phi}\cdot \vec{\phi} )^2\right],
\label{Hphi4}
\end{equation}
where $\vec{\phi}(x)$ is an  $N$-component real field.
Various computational methods, supported by
renormalization-group (RG) theory, have provided accurate determinations of
several universal quantities, see, e.g.,  Ref.~\cite{review}
for a recent comprehensive review. Among others, we should mention
the critical exponents, the equation of state, and the 
correlation functions of the order parameter $\vec{\phi}(x)$. 
However, 
for some experimental systems one is also interested in the
behavior of correlation functions describing the critical fluctuations
of secondary, quadratic local fields.
Due to the symmetry of the theory, there are 
two independent quantities that are quadratic in 
the fundamental field $\vec{\phi}(x)$:
one is the local energy density
\begin{equation}
E(x) = \vec{\phi}(x)\cdot \vec{\phi}(x),
\label{edef}
\end{equation}
which is O($N$) invariant;
the other one is  the anisotropic second-order traceless tensor
\begin{equation}
T_{ij}(x) = \phi_i(x) \phi_j(x) - \delta_{ij}\, {1\over N} \,
\vec{\phi}(x)\cdot \vec{\phi}(x).
\label{Tdef}
\end{equation}

The crossover exponent $\phi_T$ associated with 
the traceless tensor field $T_{ij}(x)$ 
describes the instability of the  O($N$)-symmetric 
theory against anisotropy \cite{FP-72,Wegner-72,FN-74,Aharony-76}. 
It is thus relevant for the description of 
multicritical phenomena, for instance the critical behavior near 
a bicritical point where two critical lines with $O(N)$ and 
$O(M)$ symmetry meet, giving rise to a critical theory with enlarged
$O(N+M)$ symmetry, see, e.g., Refs. \cite{PJF-74,Fisher-75,KNF-76}. 
This bicritical behavior has been the object of new studies quite recently,
since it appears in the $SO(5)$ theory of superconductivity 
\cite{Zhang-97}, and has been observed experimentally in 
organic conductors \cite{MN-00}. 
As discussed in Ref.~\cite{NA-97},  
the correlation functions  $G_E(x-y)\equiv \langle 
E(x) E(y) \rangle$ and 
$G_T(x-y) \equiv \langle T_{ij}(x) T_{ij}(y) \rangle$ 
are relevant in the description of 
strain-strain correlations in certain liquids and solids, 
where an effective coupling between the order parameter and the
elastic deformations occurs.
Moreover, in the special case $N=2$, the traceless 
tensor field $T_{ij}(x)$
is related to the second-harmonic order parameter
in density-wave systems, whose critical behavior 
belongs to the $XY$ universality class, see, e.g.,
 Refs.~\cite{LB-82,Fawcett-88,NA-97}.
Experimentally, such behavior
is observed at the nematic-smectic A transition in liquid crystals
\cite{LB-82,ABBL-86,Girault-etal-89,Brock-etal-89,Garland-etal-93,%
Wu-etal-94,Aharony-etal-95,NA-97}. 
In these systems the structure factor of the secondary order parameter $T_{ij}$
has been measured using x-ray scattering techniques 
\cite{Wu-etal-94,Aharony-etal-95}. 
The crossover exponent $\phi_T$ is also relevant \cite{Aharony-91}
in the description of crossover effects in diluted Ising 
antiferromagnets with $n$-fold degenerate ground state
\cite{Fernandez-88}, for instance in some diluted magnetic 
semiconductors such as Cd${}_{1-x}$Mn$_x$Te.

In this paper we determine the crossover 
exponent $\phi_T$. Such a quantity has already been obtained
in the framework of the $\epsilon$-expansion to three loops
\cite{Yamazaki-74}, from the analysis of high-temperature expansions
\cite{PJF-74} for $N=2,3$, and by means of a Monte Carlo simulation 
\cite{Hu-01} for $N=5$. Here, 
we consider the alternative field-theoretical (FT) method 
based on a fixed-dimension expansion in powers of the 
zero-momentum quartic coupling \cite{Parisi_Cargese}, and 
perform a six-loop calculation of $\phi_T$. 
For the physically interesting cases $N=2,3,5$ we obtain
\begin{eqnarray}
&& \phi_T = 1.184(12) \qquad\qquad (N = 2), \nonumber \\
&& \phi_T = 1.271(21) \qquad\qquad (N = 3), \nonumber \\
&& \phi_T = 1.40(4)\hphantom{12} \qquad\qquad   (N = 5). 
\end{eqnarray}

We also consider the correlation functions $G_E(x)$ and 
$G_T(x)$ in the high-temperature phase.
In the critical limit,
the Fourier transform $\widetilde{G}_T(q)$ obeys a scaling law that 
is analogous to that of the fundamental correlation function, i.e.
\begin{equation}
\widetilde{G}_T(q,t) = A_T^+ t^{-\gamma_T} f_T(q^2\xi^2),
\end{equation}
where $t\equiv (T-T_c)/T_c$ is the reduced temperature,
$\gamma_T = 2 \phi_T - 2 + \alpha$ is the tensor susceptibility 
exponent and $\xi$ is the second-moment correlation length computed from
the two-point function of the order parameter. 
The same scaling 
behavior holds for the correlation functions
$\widetilde{G}_E(q,t)$ of systems in  the Ising universality class, with 
$\alpha$ replacing $\gamma_T$, i.e.
$\widetilde{G}_E(q,t) = A_E^+ t^{-\alpha} f_E(q^2\xi^2)$.
For $N\ge 2$, however, $\alpha$ is negative
and an additional background term should be taken into account. In this 
case, in the critical limit, we have
\begin{equation}
\widetilde{G}_E(q,t) = B_E + \widetilde{G}_{E,\rm sing}(q,t) 
               = B_E +  A_E^+ t^{-\alpha} f_E(q^2\xi^2).
\end{equation}
The background term $B_E$ is the dominant one and the singular part
vanishes at criticality. In this case, by using 
positivity (unitarity in FT language) arguments, one may also show 
that $A_E^+<0$, as observed in experiments.

In this paper we extend the two-loop $\epsilon$-expansion 
computation of Refs.~\cite{NA-97,Aharony-etal-95}. 
We compute the universal scaling functions $f_E(q^2\xi^2)$ 
and $f_T(q^2\xi^2)$ using the $\epsilon$ expansion and the 
expansion in fixed dimension $d=3$. First, we determine the small-momentum 
behavior to four loops in the fixed-dimension expansion and to three
loops in $\epsilon$-expansion. In particular, we obtain accurate
estimates of the experimentally relevant ratios 
$X_{E,T} \equiv \xi^2_{E,T}/\xi^2$, where $\xi_{E,T}$ is the 
second-moment correlation length computed from $G_{E,T}(x)$ or from 
its singular part if $\alpha$ is negative.  
For instance, for $N=1$ we find 
\begin{equation}
X_E = 0.0140(5),
\end{equation}
and for $N=2$ 
\begin{equation}
X_E = -0.0017(1), \qquad\qquad X_T = 0.041(2).
\end{equation}
Moreover, we study the large-momentum behavior of the structure factors 
and construct interpolations valid for all momenta by using the 
dispersive approach applied to $\langle\phi(0)\phi(x)\rangle$ 
by Bray \cite{Bray-76}. 

The paper is organized as follows. 
In Sec.~\ref{sec2} we report the computation of the crossover exponent 
$\phi_T$ to six loops in the fixed-dimension expansion and compare 
our results with the existing theoretical and experimental estimates
(Sec.~\ref{sec2.D}). In Sec.~\ref{sec3} we report the computation of the 
structure factors. In Sec.~\ref{sec3.A} we briefly summarize the 
expected behavior of the structure factors in the critical region 
and set our notations. In Sec.~\ref{sec3.B} we explain our FT 
calculation, whose results are presented in Sec.~\ref{sec3.C}. 
In Sec.~\ref{sec3.D} we finally give approximate expressions for the 
structure factors by using a dispersive approach. 
Appendix~\ref{app2} discusses 
the large-momentum  behavior of the structure factors.
Details of the perturbative calculation  
are reported in App.~\ref{diag}.

\section{The crossover exponent associated with the tensor composite field}
\label{sec2}

\subsection{Zero-momentum scaling behavior}
\label{sec2.A}

The zero-momentum behavior of correlation functions involving
generic local operators ${\cal O}(x)$,
such as $E(x)$ and $T_{ij}(x)$, can be obtained from the 
free energy in the presence of an
external field $h_{\cal O}$ coupled with ${\cal O}(x)$. Indeed,
the singular part of the free energy scales as \cite{PJF-74}
\begin{equation}
F_{\rm sing} \propto t^{2-\alpha} f\left( h/t^{\beta+\gamma},
  h_{\cal O}/t^{\phi_{\cal O}} \right),
\label{freeen}
\end{equation}
where $h$ is the magnetic field,
and $\phi_{\cal O}$ is the crossover exponent. 
Then, by differentiating with respect to $h_{\cal O}$, one obtains
the zero-momentum correlations and the RG relations 
\begin{eqnarray}
\beta_{\cal O} &=& 2 - \alpha  - \phi_{\cal O},
\nonumber \\
\gamma_{\cal O} &=& - 2 + \alpha + 2\phi_{\cal O},
\label{scalrel}
\end{eqnarray}
where the exponents $\beta_{\cal O}$ and  $\gamma_{\cal O}$ describe 
respectively the critical (singular) behavior of 
the average $\langle {\cal O}(x) \rangle \sim |t|^{\beta_{\cal O}}$ and 
of the susceptibility
$\chi_{\cal O} \equiv \sum_x \langle {\cal O}(0) {\cal O}(x) \rangle_c 
\sim t^{-\gamma_{\cal O}}$.

In this section  we compute the
crossover exponent $\phi_T$ associated with
the tensor field $T_{ij}(x)$ in the fixed-dimension
FT framework, by performing a six-loop
perturbative expansion.
Of course, the crossover exponent associated
with the energy density $E(x)$ is trivial, i.e. $\phi_E=1$
and $\gamma_E=\alpha$.

\subsection{The fixed-dimension expansion: generalities}
\label{sec2.B}

In the fixed-dimension FT approach, one 
renormalizes the theory by introducing a set of zero-momentum conditions 
for the two-point and four-point one-particle irreducible correlation functions
\begin{eqnarray}
&& \hskip -1truecm
\Gamma^{(2)}_{ij}(p) = \delta_{ij} Z_\phi^{-1} \left[ m^2+p^2+O(p^4)\right],
\label{ren1g}  \\
&& \hskip -1truecm
\Gamma^{(4)}_{ijkl}(0) = m^\epsilon\,Z_\phi^{-2} \,g \,\case{1}{3} \left(
\delta_{ij}\delta_{kl} +  \delta_{ik}\delta_{jl} +  \delta_{il}\delta_{jk} 
\right),
\label{rencond}  
\end{eqnarray} 
where $\epsilon\equiv 4 - d$ and $d$ is the space dimension.
They relate the mass $m$ and the zero-momentum
renormalized coupling $g$ to the corresponding Hamiltonian parameters
$r$ and $u$:
\begin{equation}
u = m^\epsilon \,g \, Z_u(g) \, Z_\phi(g)^{-2}.
\label{ug}
\end{equation}
In addition, one introduces the function $Z_t$ that is defined by the relation
\begin{equation}
\Gamma^{(1,2)}_{ij}(0) = \delta_{ij} Z_t(g)^{-1},
\label{rencond2}
\end{equation}
where $\Gamma^{(1,2)}(p)$ is the one-particle irreducible
two-point function with an insertion of $\case{1}{2} \vec{\phi}^{\;2}$.

The critical theory is obtained by setting $g=g^*$, where 
$g^*$ is the nontrivial zero of the $\beta$-function
\begin{equation}
\beta(g) =  m \left. {\partial g \over \partial m}\right|_{u} .
\label{beta}
\end{equation}
The standard critical exponents are then obtained by evaluating 
the RG functions
\begin{eqnarray}
\eta_\phi(g) &=& \left. {\partial \ln Z_\phi\over \partial \ln m}
      \right|_{u} , 
\nonumber \\
\eta_t(g) &=& \left. {\partial \ln Z_t \over \partial \ln m} 
      \right|_u
\label{defesponentiphi-t}
\end{eqnarray}
at the fixed point $g^*$, i.e. 
\begin{eqnarray}
\eta &=& \eta_\phi(g^*),\nonumber \\
{1\over \nu}  &=& 2 + \eta_t(g^*) - \eta_\phi(g^*) \, .
\label{etanu}
\end{eqnarray}
In three dimensions these RG functions
are known to six loops for generic values of $N$ 
\cite{BNGM-77,AS-95}. For $N=0,1,2,3$, seven-loop series for 
$\eta_\phi$ and $\eta_t$ were computed in Ref. \cite{MN-91}. 

In order to evaluate the crossover exponent $\phi_T$ associated with the 
operator $T_{ij}(x)$, we define the
renormalization function $Z_T(g)$ from the one-particle irreducible
two-point function $\Gamma_T^{(2)}(p)$ with an insertion of the operator
$T_{ij}$, i.e. we set
\begin{equation}
\Gamma_T^{(2)}(0)_{ij;k,l} = Z_T^{-1}(g) \; A_{ijkl},
\end{equation}
where
\begin{equation}
A_{ijkl} = \delta_{ik}\delta_{jl} + \delta_{il}\delta_{jk} - 
{2\over N}  \delta_{ij} \delta_{kl},
\end{equation}
so that $Z_T(0)=1$.
Then, we compute  the RG function 
\begin{equation}
\eta_T(g) = \left. {\partial \ln Z_T \over \partial \ln m} \right|_u 
= \beta(g) {d \ln Z_T \over d g}, 
\label{defesponenteetaT}
\end{equation}
and $\eta_T = \eta_T(g^*)$.
Finally, the RG scaling relation
\begin{equation}
\phi_T = \left( 2 + \eta_T - \eta\right)\nu
\label{scalrel2}
\end{equation}
allows us to determine $\phi_T$.

\subsection{The fixed-dimension expansion: six-loop results}
\label{sec2.C}

We computed $\Gamma^{(2)}_T(0)$ to six loops. 
The calculation is rather cumbersome, since it requires
the evaluation of 563 Feynman  diagrams.
We handled it with a symbolic manipulation program, which  generates the diagrams 
and computes the symmetry and group factors of each of them.
We used the numerical results compiled in Ref.~\cite{NMB-77}
for the integrals associated with each diagram.
We obtained
\begin{eqnarray}
&&\eta_T(\bar{g}) =  - \bar{g} {2\over 8+N} + \bar{g}^2 {2 (6+N)\over
3(8+N)^2}  
- \bar{g}^3 { 18.312844 + 3.433275 N - 0.21674589 N^2 \over (8+N)^3}
 \\
&& + \bar{g}^4 {140.79937 + 37.573408 N + 1.0362736 N^2  + 0.094342565
N^3\over (8+N)^4}  \nonumber \\
&&- \bar{g}^5 { 1340.075 + 416.71657 N +  17.622623 N^2  - 0.91128056 N^3
- 0.050833747 N^4 \over (8+N)^5 } \nonumber \\
&&+ \bar{g}^6 {15651.266 + 5665.6519 N + 433.68712 N^2  + 1.0675503 N^3  + 
    0.67910559 N^4  + 0.031393004 N^5 \over (8+N)^6} \nonumber\\
&& + O(\bar{g}^7), \nonumber
\end{eqnarray}
where, as usual, we have introduced the rescaled coupling $\bar{g}$
defined by 
\begin{equation}\label{gnew}
g  =  {48 \pi\over 8+N} \;\bar{g} .
\end{equation}
Field-theoretical  perturbative expansions are divergent, and thus,
in order to obtain accurate results, an appropriate resummation 
is required. We use the method of Ref.~\cite{LZ-77} that takes into account 
the large-order behavior of the perturbative expansion, 
see, e.g., Ref.~\cite{Zinn-Justin-book}. Mean values and error
bars are computed using the algorithm of Ref. \cite{CPV-00}.

Given the expansion of $\eta_T(\bar{g})$, we determine the perturbative 
expansion of $\phi_T(\bar{g})$, $\beta_T(\bar{g})$, and 
$\gamma_T(\bar{g})$, using the relations (\ref{scalrel}) and 
(\ref{scalrel2}). For $N=2$ we obtain \cite{foot1}
$\phi_T = 1.176(4)$, 1.178(3), $\beta_T = 0.821(6)$, 0.825(5), and 
$\gamma_T = 0.355(2)$, 0.358(3), where, for each exponent, 
we report the estimate obtained from the direct analysis 
and from the analysis of the series of the inverse, i.e.
from $1/\phi_T(g)$, etc. The two estimates obtained for each exponent
agree within error bars, but, with the quoted errors, 
the scaling relations (\ref{scalrel}) are not well satisfied.
For instance, using $\nu = 0.67155(27)$ (Ref. \cite{CHPRV-01})
and $\beta_T = 0.823(6)$ we obtain $\phi_T = 1.192(6)$, while using 
the same value of $\nu$ and $\gamma_T = 0.3565(30)$ we have 
$\phi_T = 1.1855(15)$. These two estimates are slightly higher 
than those obtained from the analysis of $\phi_T(g)$ and $1/\phi_T(g)$. 
Clearly, the errors are somewhat underestimated, a phenomenon
that is probably connected with the nonanalyticity
\cite{PV-98,CCCPV-00,CPV-01} of the 
RG functions at the fixed point $\bar{g}^*$.

In order to obtain a conservative estimate, we have thus decided to 
take as estimate of $\phi_T$ the weighted average of the direct 
estimates and of the estimates obtained using $\beta_T$ and $\gamma_T$
together with the scaling relations \cite{foot2}.
The (very conservative) error is such to include all estimates. 
The other exponents are dealt with analogously. 
The final results for several values of $N$ are reported in 
Table~\ref{phiTres}.

\begin{table}[tbp]
\caption{
Critical exponents associated with the tensor field $T_{ij}(x)$.
}
\label{phiTres}
\begin{tabular}{clll}
\multicolumn{1}{c}{$N$}&
\multicolumn{1}{c}{$\phi_T$}&
\multicolumn{1}{c}{$\beta_T$}&
\multicolumn{1}{c}{$\gamma_T$}\\
\tableline \hline
2   &   1.184(12)   &   0.830(12)   &   0.354(25) \\
3   &   1.271(21)   &   0.863(21)   &   0.41(4)   \\  
4   &   1.35(4)     &   0.90(4)     &   0.45(8)   \\
5   &   1.40(4)     &   0.90(4)     &   0.50(8)   \\
8   &   1.55(4)     &   0.94(4)     &   0.61(8)   \\
16  &   1.75(6)     &   0.98(6)     &   0.77(12)  \\
\end{tabular}
\end{table}

\subsection{Comparison with previous results} \label{sec2.D}

The exponent $\phi_T$ can also be computed in the 
$\epsilon$ expansion. Three-loop series were derived in 
Ref.~\cite{Yamazaki-74}:
\begin{eqnarray}
\phi_T = && 1 + \epsilon {N\over 2(N+8)}  + 
\epsilon^2 {N^3+24N^2+68N\over 4 (N+8)^3}  \label{phitexp} \\
&&+ \epsilon^3 { N^5 +48N^4+788N^3+3472N^2 + 5024N - 48N(5N+22)(N+8) 
\zeta(3) \over 8 (N+8)^5}  + O(\epsilon^4).
\nonumber
\end{eqnarray}
The coefficients of this series decrease rapidly; for instance, we have
\begin{equation}
\phi_T(N=2) = 1 + 0.1 \epsilon + 0.06 \epsilon^2 - 0.00735899 \epsilon^3+ 
O(\epsilon^4), 
\end{equation}
\begin{equation}
\phi_T(N=3) = 1 + 0.136364 \epsilon + 0.083959 \epsilon^2  + 
  0.000991 \epsilon^3 + O(\epsilon^4),
\end{equation}
for $N=2$ and 3 respectively. Thus, any resummation gives estimates that 
do not differ significantly from those obtained by simply setting 
$\epsilon = 1$. For $N=2,3$ we obtain 
$\phi_T \approx 1.15$, $\phi_T \approx 1.22$, in reasonable  agreement---keeping 
into account that these are three-loop results---with the 
estimates of Table \ref{phiTres}.
They are also in agreement with the estimate of 
Ref. \cite{ABBL-86} that reports $\phi_T=1.16(7)$ for $N=2$,
which has been obtained by analyzing the same $O(\epsilon^3)$ series
and the two-loop series calculated in the framework of 
the fixed-dimension expansion.

The exponent $\phi_T$ has been also computed in the $1/N$ 
expansion \cite{largen} for $d=3$:
\begin{equation}
\phi_T = 2 -  {32\over \pi^2 N} + O\left( {1\over N^2} \right).
\end{equation}
For $N=16$ it gives $\phi_T = 1.80$, which agrees with  the 
FT result of Table \ref{phiTres}.

The exponent
$\phi_T$ has been estimated by high-temperature expansion techniques
in Ref.~\cite{PJF-74}, obtaining $\phi_T=1.175(15)$ for $N=2$ and
$\phi_T=1.250(15)$ for $N=3$, in agreement with the FT estimates.
For $N=5$, the exponent $\phi_T$ has also been determined 
by means of a Monte Carlo simulation \cite{Hu-01}: 
$\phi_T = 1.387(30)$.

Experimental estimates of $\phi_T$ are
reported in Ref.~\cite{PHA-91}. We mention the experimental
result $\phi_T=1.17(2)$ for the ($2\rightarrow 1 + 1$) bicritical point
in GdAlO$_3$ \cite{RG-77}.
The  ($3\rightarrow 2 + 1$) bicritical behavior has been 
studied in MnF$_2$ \cite{KR-79}, obtaining $\phi_T=1.279(31)$.
The experimental results obtained for a nematic--smectic-A transition
reported in Ref.~\cite{Wu-etal-94} are $\beta_T=0.76(4)$ and
$\gamma_T = 0.41(9)$.

\section{The structure factor of the bilinear fields in the high-temperature 
         phase}
\label{sec3}

\subsection{Scaling behavior}
\label{sec3.A}

The two-point correlation function of the fundamental field,
i.e. $G(x)=\langle \vec{\phi}(0)\cdot \vec{\phi}(x)\rangle $, 
is of  central importance because its Fourier transform
$\widetilde{G}(q)$ is directly related to the scattering intensity 
in scattering experiments.  For $t\rightarrow 0^+$,
its asymptotic behavior is given by \cite{BF-67,FA-73} 
\begin{equation}
\widetilde{G}(q) = C^+ t^{-\gamma} f(q^2\xi^2),
\label{fg}
\end{equation}
where $C^+$ is the amplitude of the magnetic susceptibility and 
the function $f(y)$ is universal.
Taking the second-moment correlation length 
\begin{equation}
\xi^2 \equiv {1\over 2d}\, {\sum_x |x|^2 G(x)\over 
                     \sum_x G(x)} = -\widetilde{G}(0)^{-1} 
\left. {\partial \widetilde{G}(q) \over \partial q^2}\right|_{q^2=0}
\label{xism}
\end{equation}
as length scale, 
the small-momentum behavior of $f(y)$ is
$f(y) = 1/(1 + y) + O(y^2)$, with very small $O(y^2)$ corrections.
Theoretical results for the correlation function $\widetilde{G}(q)$ are 
reviewed, e.g., in Ref.~\cite{review}.

In this section we study the scaling behavior of  the 
two-point correlation functions of the 
bilinear fields $E(x)$ and $T_{ij}(x)$. 
Like the specific heat, which is given by the
zero-momentum component of the two-point function $\widetilde{G}_E(q,t)$,
the asymptotic behavior of $\widetilde{G}_{E,T}(q,t)$
for $t\rightarrow 0^+$ is not as simple as that of the fundamental
two-point function. 
Indeed, in the scaling limit $t\rightarrow 0^+$, $q^2\to 0$ with 
$q^2\xi^2$ fixed, RG theory predicts
\begin{equation}
\widetilde{G}_{E,T}(q,t) = B_{E,T}
   \left[1 + O(t)\right] + A_{E,T}^+ t^{-\gamma_{E,T}} f_{E,T}(q^2\xi^2) 
   \left[1 + O(t^\Delta)\right],
\label{Go}
\end{equation}
where $B_{E,T}$ and $A_{E,T}^+$ are nonuniversal constants, 
$f_{E,T}(y)$ is a universal function satisfying $f_{E,T}(0)=1$, 
and $\Delta$ is the exponent related to the leading irrelevant operator.
As amply discussed in textbooks---see, e.g., 
Ref.~\cite{Zinn-Justin-book}---the presence of the background term 
$B_E$ in the asymptotic behavior  of $\widetilde{G}_E(q,t)$
is related to the
need of an additive renormalization.
One may easily see that the same argument applies to the two-point 
function  $\widetilde{G}_T(q,t)$ of $T_{ij}$.

Since $\gamma_T > 0$ for all $N\ge 2$,
the leading behavior of the tensor two-point function is
determined by the singular term depending on the scaling function
$f_T(q^2\xi^2)$:
\begin{equation}
\widetilde{G}_{T}(q,t) = A_{T}^+ t^{-\gamma_{T}} f_{T}(q^2\xi^2) 
\left[ 1 + O(t^\Delta) + O(t^{\gamma_T}) \right]\; .
\label{GT}
\end{equation}
The background term $B_T$ gives subleading corrections of order
$t^{\gamma_T}$, that turn out to be more relevant 
than the standard scaling corrections of order $t^\Delta$. 
Indeed, for the physically relevant cases 
$N=2,3$, one finds that  $\gamma_T < \Delta$
($\Delta\approx 0.53$ for $N=2$ and $\Delta\approx 0.55$  for $N=3$,
see, e.g., the results reviewed in Ref.~\cite{review}). 
The difference decreases as $N\to \infty$, since both 
$\gamma_T$ and $\Delta$ converge to 1 with the same $1/N$ correction. 
 
The same thing holds for the energy two-point function in the case
of the Ising universality class for which $\alpha$ is positive,
$\alpha=0.1199(7)$ (Ref. \cite{CPRV-99}), i.e.
\begin{equation}
\widetilde{G}_{E}(q,t) = A_{E}^+ t^{-\alpha} f_{E}(q^2\xi^2) 
\left[ 1 + O(t^\Delta) + O(t^\alpha) \right],
\label{GE-N1}
\end{equation} 
where $\Delta\approx 0.53$, see e.g. Ref.~\cite{Hasenbusch-99}.
On the other hand, 
for the $O(N)$ vector models with $N\ge 2$, since $\alpha<0$,
the background term $B_E$ gives 
the leading behavior of the energy two-point function 
$\widetilde{G}_E(q,t)$:
\begin{equation}
\widetilde{G}_{E}(q,t) = B_{E} + A_{E}^+ t^{-\alpha}
f_{E}(q^2\xi^2) \left[ 1 + O\left(t^{\Delta}\right) \right] + O\left(t \right).
\label{Ges}
\end{equation} 
In these cases, the singular part vanishes for $t=0$ and is usually responsible
for a cusp-like finite maximum in the specific heat 
at the critical point,  as it is observed in experiments and in
lattice models.
This requires the  nonuniversal constant $A^+_E$ to be negative
(see the discussion in Sec. \ref{sec3.C.1}).

In order to single out the singular behavior, one may consider the 
derivative with respect to the reduced temperature $t$
\begin{equation}
W_{E,T}(q,t) \equiv {\partial \widetilde{G}_{E,T}\over \partial t } =
- \gamma_{E,T} A^+_{E,T} t^{-1-\gamma_{E,T}} w_{E,T}(q^2\xi^2) 
\left[1 + O(t^\Delta, t^{1+\gamma_{E,T}})\right],
\label{defw}
\end{equation}
where
\begin{equation}
w_{E,T}(y) = f_{E,T}(y) + {2\nu\over \gamma_{E,T}} y f_{E,T}'(y) = 1 + O(y)
\label{wodef}
\end{equation}
is another universal function.

\subsubsection{Small-momentum behavior}
\label{sec3.A.1}

At small momentum, i.e. for $y \equiv q^2\xi^2\ll 1$,
the scaling functions $f_{E,T}(y)$ behave as 
\begin{eqnarray}
&&f_{E}(y) = 1 + \sum_{n=1} e_n y^n,\label{fey}\\
&&f_{T}(y) = 1 + \sum_{n=1} a_n y^n. \label{fty}
\end{eqnarray}
Using Eq.~(\ref{wodef}),
these expansions can be related to those of the scaling functions
$w_{E,T}(y)$,
\begin{eqnarray}
&&w_{E}(y) = 1 + \sum_{n=1} \bar{e}_n y^n, \label{wey}\\
&&w_{T}(y) = 1 + \sum_{n=1} \bar{a}_n y^n.\label{wty}
\end{eqnarray}
Indeed, it is immediate to obtain 
\begin{eqnarray}
&&\bar{e}_n = e_n \left( 1 + {2n\nu \over \alpha} \right), 
\label{ebn} \\
&&\bar{a}_n = a_n \left( 1 + {2n\nu \over \gamma_T} \right).
\label{abn} 
\end{eqnarray}
Simple arguments based on perturbation theory suggest
that the convergence radius $R_c$ 
of the small-momentum expansions is determined
by the two-particle cut. 
The singularity in the complex plane closest to the origin is expected to be
$y_s= - 4 S^+_M$, where $S^+_M={\xi^2/\xi_{\rm gap}^2}$ and 
$\xi_{\rm gap}$ is the exponential  correlation length that determines the
large-distance exponential behavior of the fundamental two-point function.
Therefore, $R_c=4 S^+_M$.
For the $O(N)$ vector models, $S^+_M$ is very close to
one, so that $R_c\approx 4$.
For example, $S^+_M= 0.999634(4)$ for the Ising universality class
\cite{CPRV-99},
$S^+_M=0.999592(6)$ for the $XY$ universality class \cite{CHPRV-01},
$S^+_M=0.99959(4)$ for the Heisenberg universality class
\cite{CPRV-98}, and $S^+_M = 1 - 0.004590/N + O(1/N^2)$ in the
large-$N$ limit \cite{CPRV-98}.
As a consequence, for $n\to \infty$, 
\begin{equation}
{e_{n+1}\over e_n} \approx 
{a_{n+1}\over a_n} \approx 
{\bar{e}_{n+1}\over \bar{e}_n} \approx 
{\bar{a}_{n+1}\over \bar{a}_n} \approx -{1\over 4}.
\end{equation}
The constants $e_1$ and $a_1$ are related to the 
universal ratios $X_{E,T}\equiv \xi_{E,T}^2/\xi^2$ introduced in
Refs. \cite{Aharony-etal-95,NA-97}, 
where $\xi_{E,T}$ are the second-moment correlation lengths
associated with the singular part of the energy and of the tensor 
two-point functions respectively. More precisely, if $\gamma_{T,E} > 0$, 
the correlation length is defined by Eq. (\ref{xism}), replacing 
$\widetilde{G}(q)$ with $\widetilde{G}_{E,T}(q)$. If the exponent is negative,
then
\begin{eqnarray}
\xi^2_E = - (\widetilde{G}_E(0) - B_E)^{-1}
 \left. {\partial \widetilde{G}_E(q)\over \partial q^2}\right|_{q^2=0}.
\end{eqnarray}
The universal ratios  $X_E$ and $X_T$ are given by 
$X_E= - e_1$ and $X_T = - a_1$.

\subsubsection{Large-momentum behavior}
\label{sec3.A.2}

The large-momentum behavior of the fundamental
correlation function is given by the Fisher-Langer 
formula \cite{FL-68}:
\begin{equation}
f(y) \approx {A_1\over y^{1 - \eta/2}}
  \left(1 + {A_2\over y^{(1-\alpha)/(2 \nu)}} +
            {A_3\over y^{1/(2\nu)}}\right).
\label{FL-law}
\end{equation}
One may derive a similar expression for the correlation functions
of the bilinear fields.
The large-momentum behavior of the structure factors can be
studied by performing a short-distance expansion of the
two-point functions $G_E(x)$ and $G_T(x)$. 
Following the method outlined in Ref.~\cite{BLZ-76},
we obtain the corresponding asymptotic 
expansions for $y\to \infty$:
\begin{eqnarray}
&&f_E(y) \approx  E_1 y^{-\alpha/(2\nu)} \left(
1 + {E_2 \over y^{(1-\alpha)/(2\nu)}} + 
    {E_3 \over  y^{1/(2\nu)}} \right),\label{lme}\\
&&f_T(y) \approx T_1 y^{-\gamma_T/(2\nu)} \left(
1 + {T_2 \over y^{(1 - \alpha)/(2\nu)}} +  {T_3 \over
y^{1/(2\nu)}} \right).
\label{lmt}
\end{eqnarray}
The derivation of these formulae is reported in App.~\ref{app2}.
Notice that for the $O(N)$ vector models with $N\ge 2$, since
$\alpha<0$, $f_E(y)$ increases as $y\to \infty$.

\subsection{Field-theory calculations: generalities} \label{sec3.B}

Because of the presence of the background term, the FT 
calculation of the scaling functions 
$f_{E}(y)$  and $f_{T}(y)$ requires some care. 
First, we define the dimensionless functions
\begin{equation} 
{\cal G}_{E,T}(g,y) \equiv u \widetilde{G}_{E,T}(q,t,u),
\end{equation}
where $g$ is the four-point renormalized coupling.
Then, in order to eliminate the constant additive renormalization term,
we consider the derivative with respect to $m$
of ${\cal G}_{E,T}(g,y)$:
\begin{equation}
{\cal W}_{E,T}(g,y) = 
    \left. m {\partial\over \partial m} {\cal G}_{E,T}(g,y)\right|_u = 
\beta(g) {\partial {\cal G}_{E,T}(g,y) \over \partial g} -
2 y {\partial {\cal G}_{E,T}(g,y) \over \partial y} .
\label{Hg}
\end{equation}
At the fixed point $g^*$, the functions ${\cal W}_{E,T}(g,y)$ 
differ from $W_{E,T}(q,t)$, defined in Eq.~(\ref{defw}), 
by a multiplicative factor independent of $q$.
Therefore,
the scaling functions $w_{E,T}(g,y)$, defined in Eq. (\ref{wodef}), 
are given by
\begin{equation}
w_{E,T}(g,y) = { {\cal W}_{E,T}(g,y)\over {\cal W}_{E,T}(g,0)}.
\label{wogy}
\end{equation}
Note that the zero-momentum functions ${\cal W}_{E,T}(g,0)$ are
related to  the exponents $\gamma_{E,T}$ by the relation
\begin{equation}
- {\gamma_{E,T}\over \nu} = \lim_{g\to g^*} 
    \beta(g) {d \ln {\cal W}_{E,T}(g,0)\over dg}.
\end{equation}

\subsection{Field-theoretical results}  \label{sec3.C}

\subsubsection{Small-momentum expansion} \label{sec3.C.1}

We compute the small-momentum expansion of the structure factors
to four loops in the fixed-dimension approach and to three loops 
in the $\epsilon$ expansion. 

In the fixed-dimension approach, we first determine the expansion in powers
of $g$ of the coefficients
$\bar{e}_i$ and $\bar{a}_i$ defined in Eqs.
(\ref{wey}) and (\ref{wty}). The explicit expressions are 
reported in App.~\ref{diag}.
In order to obtain numerical estimates we use the same resummation 
procedure outlined in the previous section. Our numerical 
results are presented in Tables \ref{numeb} and \ref{numab}.
Note that, as expected, the ratios 
$\bar{e}_{i+1}/\bar{e}_i$ and $\bar{a}_{i+1}/\bar{a}_i$ quickly
approach $-1/4$. 
The corresponding coefficients $e_i$ and $a_i$
are  obtained  by using the relations (\ref{ebn}) and (\ref{abn}). 
For the exponent $\nu$ we use the same values reported before 
\cite{foot2}, while for $\gamma_T$ we use the results of Table \ref{phiTres}.
In the case of $a_i$ a large part of the uncertainty is due to
the error in the exponent $\gamma_T$ that enters the relation
between  $\bar{a}_i$ and $a_i$.
The results  are reported in Table \ref{numea}.
We also performed direct analyses  of the coefficients
$e_i$, $a_i$, considering the $g$-series that can be obtained
from Eqs.~(\ref{ebn}) and (\ref{abn}). The results are 
substantially consistent with those obtained by first 
estimating $\bar{e}_i$ and $\bar{a}_i$. In the case
of $a_i$ they turn out to be more precise; we show
also them in Table \ref{numea} (third column of results).

\begin{table}[tbp]
\caption{Estimates of the coefficients $\bar{e}_{i}$
for  several values of $N$.}
\label{numeb}
\begin{tabular}{clllll}
\multicolumn{1}{c}{$N$}&
\multicolumn{1}{c}{$\bar{e}_1$}&
\multicolumn{1}{c}{$\bar{e}_2/\bar{e}_1$}&
\multicolumn{1}{c}{$\bar{e}_3/\bar{e}_2$}&
\multicolumn{1}{c}{$\bar{e}_4/\bar{e}_3$}&
\multicolumn{1}{c}{$\bar{e}_5/\bar{e}_4$}\\
\tableline \hline
   1 & $-$0.170(5) & $-$0.206(1) & $-$0.221(1)  & $-$0.229(1) & 
       $-$0.234(1) \\
   2 & $-$0.155(5) & $-$0.199(1) & $-$0.216(2) & $-$0.226(2) & 
       $-$0.232(2)\\
   3 & $-$0.142(5) & $-$0.193(2) & $-$0.213(2) & $-$0.222(2) & 
       $-$0.230(3)\\
   4 & $-$0.133(6) & $-$0.189(3) & $-$0.211(2) & $-$0.222(3) & 
       $-$0.228(3)\\
   5 & $-$0.126(6) & $-$0.186(3) & $-$0.209(3) & $-$0.221(3) & 
       $-$0.228(4)\\
   8 & $-$0.111(5) & $-$0.180(3) & $-$0.206(3) & $-$0.219(4) & 
       $-$0.227(4)\\
\end{tabular}
\end{table}

\begin{table}[tbp]
\caption{Estimates of the coefficients $\bar{a}_i$
for  several values of $N$.}
\label{numab}
\begin{tabular}{clllll}
\multicolumn{1}{c}{$N$}&
\multicolumn{1}{c}{$\bar{a}_1$}&
\multicolumn{1}{c}{$\bar{a}_2/\bar{a}_1$}&
\multicolumn{1}{c}{$\bar{a}_3/\bar{a}_2$}&
\multicolumn{1}{c}{$\bar{a}_4/\bar{a}_3$}&
\multicolumn{1}{c}{$\bar{a}_5/\bar{a}_4$}\\
\tableline \hline
   2 & $-$0.203(2)  & $-$0.224(2) & $-$0.232(1) & $-$0.236(1) & 
       $-$0.239(1) \\
   3 & $-$0.208(2) & $-$0.226(1)  & $-$0.234(1) & $-$0.238(1) & 
       $-$0.240(1) \\
   4 & $-$0.213(2) & $-$0.228(1)  & $-$0.235(1) & $-$0.239(1) & 
       $-$0.241(1) \\
   5 & $-$0.216(1) & $-$0.230(1)  & $-$0.236(1) & $-$0.240(1) & 
       $-$0.242(1) \\
   8 & $-$0.224(1)  & $-$0.235(1)  & $-$0.239(1) & $-$0.242(1) & 
       $-$0.244(1) \\
\end{tabular}
\end{table}

\begin{table}[tbp]
\squeezetable
\caption{Results of the coefficients $e_i$ and $a_i$ 
for  several values of $N$ and from various analyses: 
(a) $(d=3$) by using the fixed-dimension results for $\bar{e}_i$ and $\bar{a}_i$, 
and by directly analyzing the series for $a_i$;  
(b) $(\epsilon$-exp) by resummation of 
the three-loop $\epsilon$-expansion.}
\label{numea}
\begin{tabular}{ccccccc}
\multicolumn{1}{c}{$N$}&
\multicolumn{1}{c}{$i$}&
\multicolumn{1}{c}{$e_i$ $(d=3)$ $_{{\rm from} \; \bar{e}_i}$}&
\multicolumn{1}{c}{$a_i$ $(d=3)$ $_{{\rm from} \; \bar{a}_i}$}&
\multicolumn{1}{c}{$a_i$ $(d=3)$ $_{\rm direct}$}&
\multicolumn{1}{c}{$e_i$ $(\epsilon$-exp)}&
\multicolumn{1}{c}{$a_i$ $(\epsilon$-exp)}\\
\tableline \hline
1 & 1  &   $-$0.0137(2) & &   & $-$0.0145(11) & \\
  & 2  &       0.147(3)$\times 10^{-2}$ & &   &    
               0.17(2)$\times 10^{-2}$& \\
  &  3  &   $-$0.219(4)$\times 10^{-3}$  & &  & 
            $-$0.26(3)$\times 10^{-3}$& \\
  &  4  &      0.38(1)$\times 10^{-4}$ &  & &    
               0.47(6)$\times 10^{-4}$& \\
  &  5  &   $-$0.71(1)$\times 10^{-5}$ &  & & 
            $-$0.9(1)$\times 10^{-5}$& \\
\hline
2 &  1  &      0.0017(1) &  $-$0.042(2) & $-$0.042(1) & 0.000(2) & $-$0.041(2) \\
  &  2  &   $-$0.017(1)$\times 10^{-2}$ &     0.530(3)$\times 10^{-2}$
& 0.52(1)$\times 10^{-2}$ & 0.00(3)$\times 10^{-2}$&    0.48(3)$\times 10^{-2}$ \\
  &  3  &      0.024(2)$\times 10^{-3}$ &  $-$0.85(1)$\times 10^{-3}$  &  $-$0.84(3)$\times 10^{-3}$  &  
 0.00(5)$\times 10^{-3}$   & $-$0.75(6)$\times 10^{-3}$   \\
  &  4  &   $-$0.041(2)$\times 10^{-4}$ &     1.5(2)$\times 10^{-4}$ & 1.5(1)$\times 10^{-4}$
&    0.0(1)$\times 10^{-4}$   
&    1.3(1)$\times 10^{-4}$  \\
  &  5  &      0.076(4)$\times 10^{-5}$   &  $-$3.0(2)$\times 10^{-5}$  & $-$3.0(2)$\times 10^{-5}$  & 
0.0(2)$\times 10^{-5}$  & $-$2.6(2)$\times 10^{-5}$  \\
\hline
3 &  1  &      0.015(2)  &  $-$0.047(4)   & $-$0.0465(8)   &    0.012(2) & $-$0.045(1) \\
  &  2  &   $-$0.14(2)$\times 10^{-2}$  & 0.59(5)$\times 10^{-2}$ & 0.59(1)$\times 10^{-2}$
& $-$0.13(3)$\times 10^{-2}$ &    0.54(3)$\times 10^{-2}$ \\
  &  3  &      0.19(2)$\times 10^{-3}$  &  $-$0.96(9)$\times 10^{-3}$   &   $-$0.96(2)$\times 10^{-3}$   &   
 0.19(5)$\times 10^{-3}$ & $-$0.85(5)$\times 10^{-3}$   \\
  &  4  &   $-$0.32(4)$\times 10^{-4}$  &     1.8(2)$\times 10^{-4}$ &  1.75(7)$\times 10^{-4}$
& $-$0.3(1)$\times 10^{-4}$   &
    1.5(1)$\times 10^{-4}$   \\
  &  5  &      0.6(1)$\times 10^{-5}$    &  $-$3.4(3)$\times 10^{-5}$ &  $-$3.4(2)$\times 10^{-5}$
&    0.6(2)$\times 10^{-5}$  & $-$2.9(2)$\times 10^{-5}$  \\
\hline
4 &  1  &      0.026(1)  &  $-$0.05(1)   & $-$0.0500(6)   &    0.022(2) & $-$0.049(1) \\
  &  2  &   $-$0.23(1)$\times 10^{-2}$   &     0.63(9)$\times
10^{-2}$   & 0.645(5)$\times 10^{-2}$  
& $-$0.22(3)$\times 10^{-2}$ &    0.60(2)$\times 10^{-2}$\\
  &  3  &      0.31(1)$\times 10^{-3}$  &  $-$1.0(2)$\times 10^{-3}$  &    $-$1.06(2)$\times 10^{-3}$  &    
0.33(4)$\times 10^{-3}$ & $-$0.94(4)$\times 10^{-3}$  \\
  &  4  &   $-$0.50(2)$\times 10^{-4}$  &     1.9(3)$\times 10^{-4}$ & 1.96(6)$\times 10^{-4}$
& $-$0.6(1)$\times 10^{-4}$   &
    1.7(1)$\times 10^{-4}$  \\
  &  5  &      0.91(3)$\times 10^{-5}$    &  
$-$3.7(6)$\times 10^{-5}$    &$-$3.9(2)$\times 10^{-5}$    &    1.1(2)$\times 10^{-5}$  & $-$3.3(1)$\times 10^{-5}$ \\
\hline
5 &  1  &      0.030(1)   &  $-$0.053(6)   &  $-$0.0533(4)   &    0.029(2) & $-$0.0528(3) \\
  &  2  &   $-$0.25(1)$\times 10^{-2}$   &     0.7(1)$\times 10^{-2}$  &  0.699(5)$\times 10^{-2}$  & 
$-$0.30(2)$\times 10^{-2}$ &    0.65(2)$\times 10^{-2}$\\
  &  3  &      0.34(1)$\times 10^{-3}$   &  $-$1.2(2)$\times 10^{-3}$  &    $-$1.16(2)$\times 10^{-3}$  &    
0.44(4)$\times 10^{-3}$ & $-$1.03(3)$\times 10^{-3}$  \\
  &  4  &   $-$0.55(2)$\times 10^{-4}$  &     2.1(3)$\times 10^{-4}$& 2.15(5)$\times 10^{-4}$
& $-$0.8(1)$\times 10^{-4}$   &
    1.85(5)$\times 10^{-4}$  \\
  &  5  &      1.00(3)$\times 10^{-5}$
    &  $-$4.2(6)$\times 10^{-5}$    &    $-$4.3(2)$\times 10^{-5}$    &    
1.4(2)$\times 10^{-5}$  & $-$3.6(1)$\times 10^{-5}$  \\
\hline
8 &  1  &      0.047(1)  &  $-$0.060(6)   & $-$0.0602(1)   &    0.046(2) & $-$0.061(1) \\
  &  2  &   $-$0.35(1)$\times 10^{-2}$   &     0.8(1)$\times10^{-2}$   &     0.817(5)$\times10^{-2}$   
& $-$0.43(1)$\times 10^{-2}$ &    0.77(1)$\times 10^{-2}$ \\
  &  3  &      0.45(1)$\times 10^{-3}$  &  $-$1.4(2)$\times 10^{-3}$  &    $-$1.40(2)$\times 10^{-3}$  &    
0.62(2)$\times 10^{-3}$ & $-$1.24(2)$\times 10^{-3}$ \\
  &  4  &   $-$0.72(2)$\times 10^{-4}$  &     2.6(3)$\times 10^{-4}$    &     2.58(4)$\times 10^{-4}$    
& $-$1.05(3)$\times 10^{-4}$   &    2.24(4)$\times 10^{-4}$  \\
  &  5  &      1.28(4)$\times 10^{-5}$ 
&  $-$5.1(6)$\times 10^{-5}$    &  $-$5.1(2)$\times 10^{-5}$    
&    1.95(5)$\times 10^{-5}$   & $-$4.4(1)$\times 10^{-5}$  \\
\end{tabular}
\end{table}

In $\epsilon$ expansion we directly resum the expansions of 
$e_i$ and $a_i$ reported in App.~\ref{diag}. The results are 
also reported in Table \ref{numea} and 
are in substantial agreement with the fixed-dimension results. 
For $a_1$, we also perform a constrained analysis that makes use of the 
available results for $a_1$ in two and one dimensions. Such a method was introduced in 
Ref. \cite{LZ-87} and generalized in Refs. \cite{PV-98,PV-98eqst}. 
In many instances it has provided quite accurate results for critical
quantities.
We use the estimates 
of $a_1$ in two dimensions reported in Refs. \cite{CEMPS-96,MPS-96}: 
$a_1 = -0.0812(5)$, $-0.1014(6)$, and $-0.1313(9)$ for 
$N=3,4,8$ respectively. We also make use of the 
one-dimensional result \cite{CMPS-97},
$a_1 = - (N-1)^2/(4 N^2)$. 
{}From the analysis constrained in one dimension, in three  dimensions
we obtain: 
$a_1 = -0.0397(2)$ for $N=2$ and 
$a_1 = -0.0533(3)$ for $N=5$,
{}while from the analysis constrained in two dimensions we obtain
$a_1 = - 0.0460(3)$ for $N=3$, 
$a_1 = - 0.0507(5)$ for $N=4$, 
$a_1 = - 0.0621(2)$ for $N=8$.
Constraining the analysis both in two and one dimension, we obtain:
$a_1 = - 0.0458(1)$ for $N=3$,
$a_1 = - 0.0514(6)$ for $N=4$, 
$a_1 = - 0.0625(2)$ for $N=8$.
These results are compatible with those of Table \ref{numea}.

Taking into account the above results for $e_1$ and $a_1$,
we consider as our final estimates  the numbers reported in Table \ref{finaleiai},
where we have been rather conservative in giving the errors, which include  
all the results we have obtained.

\begin{table}[tbp]
\caption{Final estimates of the coefficients $e_1$ and $a_1$.}
\label{finaleiai}
\begin{tabular}{cll}
\multicolumn{1}{c}{$N$}&
\multicolumn{1}{c}{$e_1$}&
\multicolumn{1}{c}{$a_1$}\\
\tableline \hline
1 &    $-$0.0140(5) &   \\
2 & 0.0017(1) & $-$0.041(2) \\
3 & 0.014(3) & $-$0.046(1) \\
4 & 0.024(3) & $-$0.051(2) \\
5 & 0.030(1) & $-$0.0533(5) \\
8 & 0.047(1) & $-$0.062(2) 
\end{tabular}
\end{table}

As already noted in Ref. \cite{NA-97}, 
$e_1 = -\alpha/(6\gamma) + O(\epsilon^3)$ and 
$a_1 = -\gamma_T/(6\gamma) + O(\epsilon^3)$. These relations are 
not satisfied to order $\epsilon^3$, see App. B. Nonetheless, they 
still provide very good approximations to $e_1$ and $a_1$. 
For instance (see Ref. \cite{review} for the estimates of the 
critical exponents):
$- {\alpha /(6\gamma)}= - 0.01377(8), 0.00185(10), 0.0159(8)$ 
respectively for  $N=1,2,3$, where the error is  related to the
uncertainty on the estimates of $\alpha$ and $\gamma$. 

The coefficient $e_1$ has also been computed for $N=1$ by Monte Carlo 
simulations \cite{NG-97}. The numerical data are well described by 
$-\alpha_{\rm eff}(t)/(6\gamma_{\rm eff}(t))$, where $\alpha_{\rm eff}(t)$ 
and $\gamma_{\rm eff}(t)$ are effective exponents determined from the 
specific heat and the susceptibility.

It is interesting to note that the signs of $a_i$ and $e_i$ are 
strictly related to the signs of the amplitudes 
$A^+_{E,T}$ and of the exponents $\alpha$ and $\gamma_T$. 
First, we observe that 
in the critical limit the correlation functions are nonnegative,
i.e. $G_{E,T}(x)\ge 0$.  Indeed, the lattice $\phi^4$ model with 
nearest-neighbor couplings is exactly reflection positive  
and therefore, the above-reported inequalities are rigorously true 
for any value of the couplings. At criticality, they should hold for any model
in the same universality class. Therefore, all moments are positive, i.e.
$\sum_x |x|^{2n} G_{E,T}(x) \ge 0$. If the correlation functions have 
the scaling forms (\ref{GT}) and (\ref{GE-N1}), this implies
\begin{equation}
A^+_{E,T} \ge 0, \qquad\qquad (-1)^n a_n \ge 0.
\label{ineqA}
\end{equation}
For $N\ge 2$, using Eq. (\ref{Ges}), we obtain
\begin{equation} 
B_E \ge 0, \qquad\qquad (-1)^n A^+_E e_n \ge 0.
\label{ineqB}
\end{equation}
Relations (\ref{ineqA}) are satisfied by our results, while 
Eq. (\ref{ineqB}) and our result $e_1> 0$ imply 
$A_E^+ < 0$.
Thus, although $A^+_{E,T}$ is nonuniversal, the positivity
(unitarity in FT language) of the theory fixes its sign.

As a final remark, note that $a_1$ and $e_1$ are very small.
The unexpectedly small value of $a_1$  was crucial to provide
an explanation \cite{Aharony-etal-95,NA-97}, consistently with 
RG theory,
of the experimental results of Ref.~\cite{Wu-etal-94} 
for density-wave systems, at the nematic--smectic-A 
transition of liquid crystals. 
The values of $a_1$ are quite smaller than
what would be naively expected.
The nearest singularity in the complex $y$ plane corresponds to the
two-particle cut, thus at  large distance 
$G_T(x)\sim |x|^q \exp ( - |x|/\xi_T')$ with $\xi'_T = \xi_{\rm gap}/2$,
where $\xi_{\rm gap}$ is the exponential  correlation length that determines the
large-distance exponential behavior of the fundamental two-point function.
Positivity of $G_T(x)$ would then require that the second-moment correlation
length $\xi_T$ be smaller than $\xi'_T$. As a consequence,
since $\xi_{\rm gap}\simeq \xi$ (see Sec.~\ref{sec3.A.1}),
$a_1 \lesssim 1/4$. But this bound turns out to be
much larger than the actual value of $a_1$.

\subsubsection{Large-momentum expansion} \label{sec3.C.2}

We also compute the constants $E_i$ and $T_i$ 
of  the large-momentum behavior of $f_{E,T}(y)$.
Matching the large-momentum expansion of the two-loop 
expression of $\widetilde{G}_{E,T}(q,t)$ with Eqs.~(\ref{lme}) and (\ref{lmt}),
we obtain
\begin{eqnarray}
&&E_1 = 1 + {4-N\over 8+N} \epsilon + O(\epsilon^2),\label{E123}\\
&&E_2 = -2 + {2 (5+N)\over 8+N} \epsilon + E_{22} \epsilon^2 + 
       O(\epsilon^3),\nonumber\\
&&E_3 = 2 - {14+N\over 8+N} \epsilon + E_{32} \epsilon^2 + 
       O(\epsilon^3),\nonumber
\end{eqnarray}
and
\begin{eqnarray}
&&T_1 = 1 + {4+N\over 8+N} \epsilon + O(\epsilon^2),\label{T123}\\
&&T_2 = -{2 (4+N)\over 4-N}  + {2(4+N)(20-13N-N^2)\over (4-N)^2(8+N)}
\epsilon + T_{23} \epsilon^2 + O(\epsilon^3), \nonumber \\
&&T_3 = {2 (4+N)\over 4-N}  - {(4+N)(56-34N-N^2)\over (4-N)^2(8+N)}
\epsilon + T_{33} \epsilon^2 + O(\epsilon^3)
\nonumber.
\end{eqnarray}
Moreover, 
\begin{eqnarray}
E_{22} + E_{32} &=& - {(N^3-14N^2-140N-432)\over 2(N+8)^3} + 
             {N-4\over 12(N+8)} \pi^2, \nonumber \\
T_{22} + T_{32} &=& - {(N+4)(N^2+14N+108)\over 2(N+8)^3} - 
             {N+4\over 12(N+8)} \pi^2.
\end{eqnarray}
The constants $E_1$ and $T_1$ are in 
agreement with the results of Ref.~\cite{NA-97}.
The divergence of the coefficients $T_2$ and $T_3$ for $N\rightarrow
4$ is related to
the vanishing of the $O(\epsilon)$ term in the expansion
of  $\alpha$ \cite{T23}.

The large size of the coefficients makes it difficult to resum the 
perturbative series. For the physically interesting case of $N=2$ 
we report the result obtained by setting $\epsilon = 1$ and 
give as error the size of the last coefficient. In this way
we obtain: $E_1=1.3(3)$, $E_2=-0.7(1.3)$, and $E_3=0.3(1.7)$ for $N=1$,
and  $E_1 = 1.2(2)$, $E_2 = -0.6(1.4)$,  $E_3 = 0.4(1.6)$,
$T_1 = 1.6(6)$, $T_2 = -9(3)$, $T_3 = 8.4(2.4)$ for $N=2$.
Moreover, $E_2 + E_3 = 0.0(2)$ and $T_2 + T_3 = -0.7(1)$ for $N=2$.

\subsection{Interpolations of the structure factors} \label{sec3.D}

In Ref. \cite{NA-97} the authors discuss several approximate forms 
for $f_{E,T}(y)$. They present generalizations of the Fisher-Burford 
\cite{BF-67}
approximant for $\langle\phi\phi\rangle$. These approximations are quite crude 
and do not reproduce the full Fisher-Langer behavior for large $y$. 
A better approach based on dispersion theory was put forward by Bray 
\cite{Bray-76}. Here, we will apply the same method to the 
universal functions $f_E(y)$ and $f_T(y)$. 

A generalization of the arguments presented in Ref. \cite{Bray-76}
gives the following representation for $f_T(y)$
\begin{equation}
f_T(y) = 1 - {y T_1\over \pi} \sin\left({\pi\gamma_T\over 2\nu}\right)
\int_{4 S^+_M}^\infty dx\, {x^{-1-\gamma_T/(2\nu)}\over x + y} F_T(x),
\label{fT-dispersive}
\end{equation}
where $F_T(x)$ is the spectral function satisfying $F_T(\infty) = 1$.
We assume here that the only singularities
of $f_T(y)$ in the complex plane are branch cuts on the 
negative real axis and that the leading one corresponds to the 
two-particle state, so that the disk $|y| < 4 S_M^+$ is free of singularities.
Under this assumption, the representation (\ref{fT-dispersive}) is exact.

For generic $F_T(x)$, Eq. (\ref{fT-dispersive}) does not give the correct 
Fisher-Langer behavior (\ref{lmt}). Indeed, for $y\to \infty$ we obtain 
$f_T(y) \approx {\rm constant} + T_1 y^{-\gamma_T/(2 \nu)}$. 
We must thus require the constant to be zero. This gives the  sum rule 
\begin{equation}
{T_1\over \pi} \sin\left({\pi\gamma_T\over 2\nu}\right)
\int_{4 S^+_M}^\infty dx\, x^{-1-\gamma_T/(2\nu)}\ F_T(x)=\, 1,
\label{sumrul-pos}
\end{equation}
which allows the determination of $T_1$ once $F_T(x)$ is given.

Eq. (\ref{fT-dispersive}) applies also to $f_E(y)$ with the 
obvious replacements. However, the sum rule (\ref{sumrul-pos}) requires 
$\alpha > 0$ and can thus be used only in the Ising case. 
For $\alpha < 0$, Eq. (\ref{sumrul-pos}) is replaced by
\begin{equation}
{E_1\over \pi} \sin\left({\pi\alpha\over 2\nu}\right)
\left[{2\nu\over \alpha} \left(4 S_M^+\right)^{-\alpha/(2\nu)} + 
\int_{4 S^+_M}^\infty dx\, x^{-1-\alpha/(2\nu)}\ (F_E(x)-1)\right]=\, 1.
\label{sumrul-neg}
\end{equation}
In order to obtain approximate expressions for the structure factors,
we must assume a specific form for the spectral function. 
For this purpose, we assume, as in Ref. \cite{Bray-76}, 
that $F_T(x)$ gives the exact Fisher-Langer behavior on the cut.
Explicitly, we consider
\begin{eqnarray}
F_T(x) &=& 1 + T_2 x^{-(1-\alpha)/(2\nu)} 
   \left[\cos {\pi(1-\alpha)\over 2\nu} + \sin{\pi(1-\alpha)\over 2\nu} 
         \cot {\pi\gamma_T\over 2\nu}\right] 
\nonumber \\
&& \qquad + T_3 x^{-1/(2\nu)}
   \left[\cos {\pi\over 2\nu} + \sin{\pi\over 2\nu}
         \cot {\pi\gamma_T\over 2\nu}\right] .
\label{FFL}
\end{eqnarray}
To completely determine the spectral function, we must specify the 
constants $T_2$ and $T_3$. We use here the $\epsilon$-expansion results 
of Sec.~\ref{sec3.C.2}.
These estimates are not very precise,
but the interpolation is quite insensitive on $T_2$ and $T_3$ separately.
Indeed, what really matters is their sum $T_2 + T_3$ that is 
more accurately determined. 
In order to test these interpolations, we can compare the estimates 
of $T_1$ and $a_i$---and, analogously, of $E_1$ and $e_i$---with 
those of the preceding 
sections. For $N=2$, using $T_2 = -9$ and $T_3 = 8.4$, we obtain 
$T_1 \approx 1.56$, $a_1 \approx -0.055$, $a_2 \approx 0.008$, 
which are reasonably close to the estimates reported before.
Analogously, using $E_2=-0.6$ and $E_3 = 0.4$, we obtain 
$E_1\approx 1.00$, $e_1 \approx 0.005$ and $e_2\approx -0.0007$,
again in reasonable agreement with previous results. In particular,
the fact that $|e_1| \ll |a_1|$ is correctly predicted by the approximation.
For $N=1$, using $E_2 = -2/3$, $E_3 = 1/3$, we obtain $E_1 \approx 1.20$,
$e_1 \approx -0.016$, $e_2 \approx 0.0019$, in reasonable agreement 
with what reported above.

\begin{figure}[tbp]
\hspace{-1cm}
\vspace{0cm}
\centerline{\psfig{width=12truecm,angle=-90,file=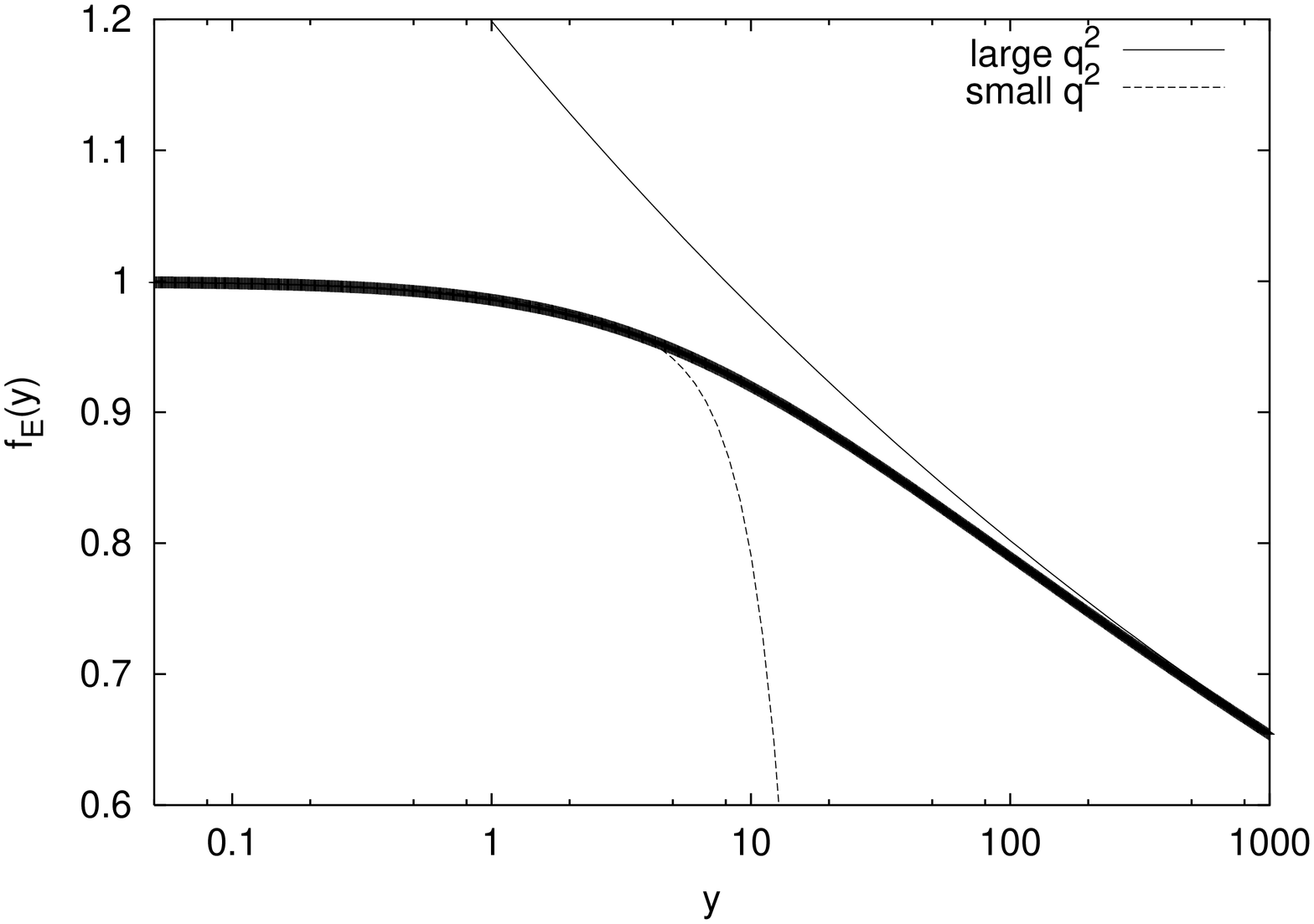}}
\vspace{0.5cm}
\caption{
Universal function $f_E(y)$ obtained using Eqs. (\ref{fT-dispersive}) 
and (\ref{FFL}), for $N=1$. We also report the large-$y$ 
behavior, $f_E(y)\approx 1.199 y^{-0.08725}$ and the small-$y$ behavior, 
$f_E(y)\approx 1 - 0.01366 y + 0.001467 y^2 - 0.000219 y^3$.
}
\label{fig:fEN1}
\end{figure}                                                                    

\begin{figure}[tbp]
\hspace{-1cm}
\vspace{0cm}
\centerline{\psfig{width=12truecm,angle=-90,file=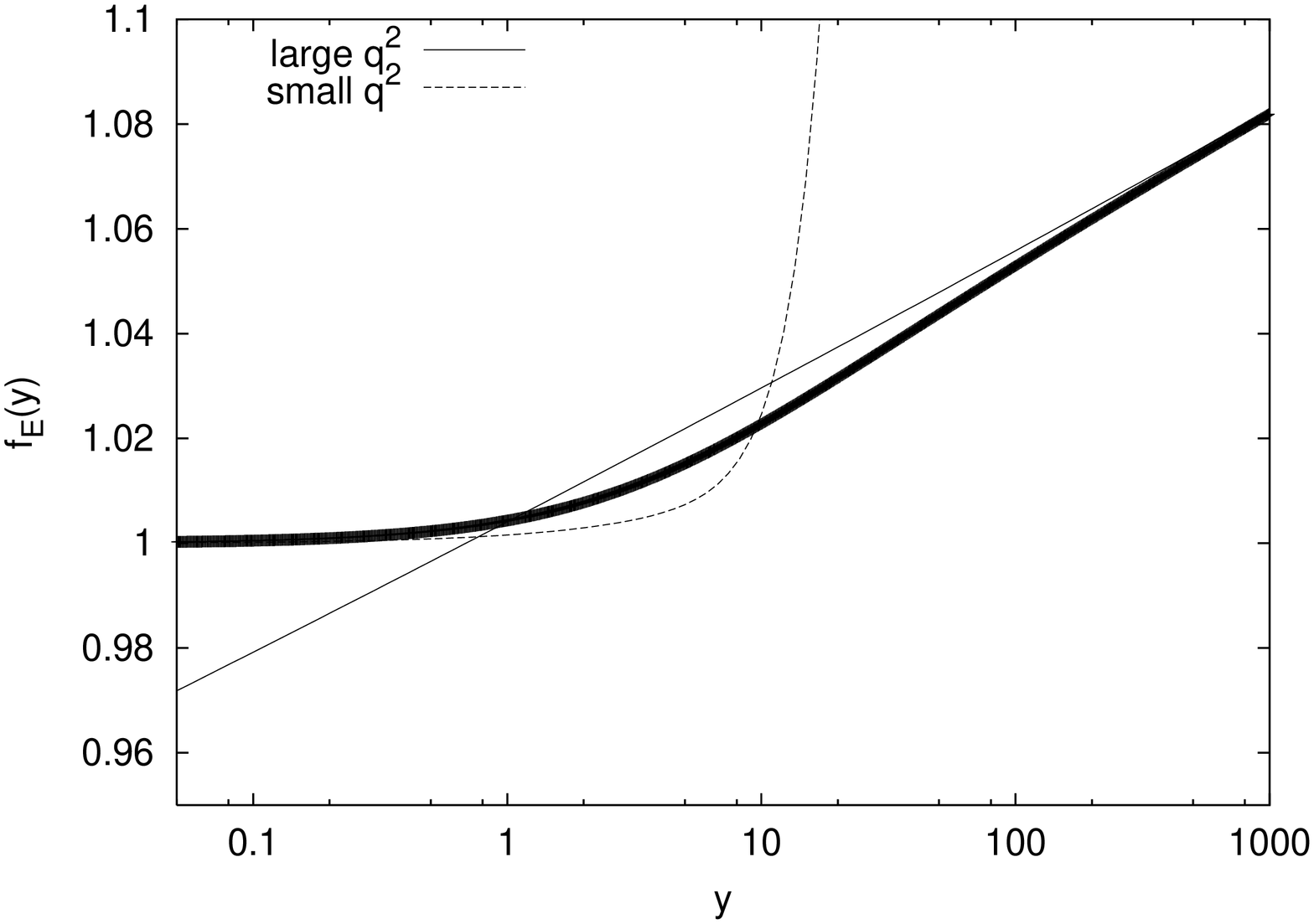}}
\vspace{0.5cm}
\caption{
Universal function $f_E(y)$ obtained using Eqs. (\ref{fT-dispersive}) 
and (\ref{FFL}), for $N=2$. We also report the large-$y$ 
behavior, $f_E(y)\approx 1.00 y^{0.010908}$ and the small-$y$ behavior, 
$f_E(y)\approx 1 + 0.00171 y - 0.000169 y^2 + 0.0000243 y^3$.
}
\label{fig:fE}
\end{figure}                                                                    

\begin{figure}[tbp]
\hspace{-1cm}
\vspace{0cm}
\centerline{\psfig{width=12truecm,angle=-90,file=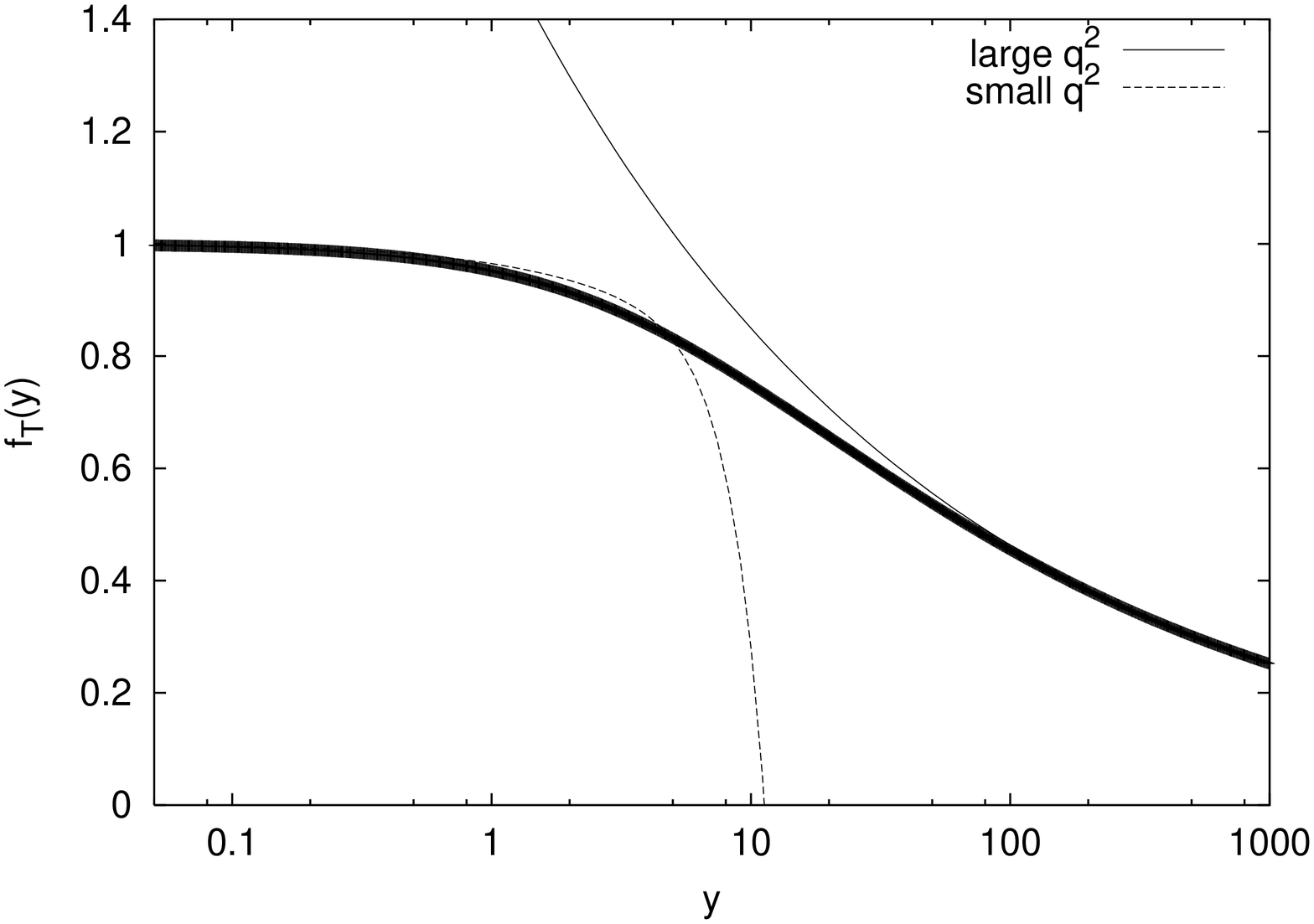}}
\vspace{0.5cm}
\caption{
Universal function $f_T(y)$ obtained using Eqs. (\ref{fT-dispersive})
and (\ref{FFL}), for $N=2$. We also report the large-$y$ 
behavior, $f_T(y)\approx 1.559 y^{-0.263569}$ and the small-$y$ behavior, 
$f_T(y)\approx 1 - 0.0397 y + 0.0053 y^2 -0.000852 y^3$.
}
\label{fig:fT}
\end{figure}                                                                    

In Fig.~\ref{fig:fEN1} we report $f_E(y)$ for $N=1$ and in 
Figs.~\ref{fig:fE} and \ref{fig:fT} a graph of $f_E(y)$ and of 
$f_T(y)$ for $N=2$. 
It is interesting to note that for $N=2$ 
the function $f_E(y)$ varies slowly and differs from one only for 
quite large values of $y$. Taking also into account that the prefactor 
vanishes as $t\to0$, the $q^2$ dependence of $\widetilde{G}_{E}(q,t)$
should be hardly visible in experiments and in numerical Monte Carlo 
simulations. Moreover, in this case $f_E(y)\ge 1$ for all $y$, 
so that, because of the inequalitites (\ref{ineqB}), there is 
an attenuation of the singular behavior
for increasing $q$, as generally expected.

\appendix

\section{Large-momentum behavior for the bilinear correlation functions}
\label{app2}

In this appendix we compute the large-momentum behavior of the 
correlation function. We follow closely the discussion of 
Refs. \cite{BAZ-74,BLZ-76} for the correlation function of the 
field $\phi$.

\subsection{The energy correlation function}

The basic ingredient of the calculation is the short-distance 
expansion of the product of operators $E(x+y/2) E(x-y/2)$.
For $y\to 0$, see, e.g., Ref.~\cite{Zinn-Justin-book},
this product is equal to the sum of all 
the operators that are allowed by  symmetries, multiplied by $C$-number
coefficients, that take in account the short-distance behavior.
The most singular contribution comes from the operators of smallest
dimension.
In this case, neglecting the 
contribution related to the identity operator, it implies
\begin{equation}
E(x+y/2) E(x-y/2)= C(y) E(x) +\mbox{less singular contributions}.
\label{SDE-EE}
\end{equation}
Now, let us consider the connected 
correlation function of $l$ composite 
operators $E(x)$, $G^{(l)}(p_1,\ldots,p_l)$, and its renormalized
counterpart $G^{(l)}_R(p_1,\ldots,p_l) = Z_t^l Z_\phi^{-l}
G^{(l)}(p_1,\ldots,p_l)$.
Then, Eq. (\ref{SDE-EE}) implies for $p \gg m$ 
\begin{equation}
G^{(l)}_R(p,-p,0,\ldots,0) \approx \widetilde{C}(p;m) G^{(l-1)}_R(0,\ldots,0),
\label{large-GR}
\end{equation}
where we have explicitly written the mass dependence of the short-distance 
coefficient. Since renormalized correlation functions scale canonically, 
i.e.
\begin{equation}
G^{(l)}_R(p,-p,0,\ldots,0) = m^{d-2l} f(p/m),
\end{equation}
we have
\begin{equation}
\widetilde{C}(p;m) = m^{-2} \hat{C}(p/m)
\label{scaling-Ctilde}
\end{equation}
Renormalized correlation functions satisfy the Callan-Symanzik equation
\begin{equation}
\left[m{\partial\over \partial m} + 
  \beta(g) {\partial\over \partial g} - l \eta_2(g)\right]
      G^{(l)}_R(p_1,\ldots,p_l) =\, 
  m^2 \sigma(g) G^{(l+1)}_R(0,p_1,\ldots,p_l),
\end{equation}
where $\sigma(g)$ is a RG function satisfying $\sigma(g^*) = 2 - \eta$, 
and $\eta_2(g) = \eta_t(g) - \eta_\phi(g)$. Applying the 
Callan-Symanzik equation to the relation (\ref{large-GR}) we obtain, setting 
$g=g^*$, 
\begin{equation}
\left[m{\partial\over \partial m} - \eta_2(g^*)\right]\widetilde{C}(p;m) = 0,
\end{equation}
and therefore, using  Eq. (\ref{scaling-Ctilde}), we have
\begin{equation}
\widetilde{C}(p;m) \sim m^{-2} (p/m)^{-2-\eta_2} = m^{-2} (p/m)^{-1/\nu}.
\end{equation}
Now, using the above-reported results and 
$Z_t/Z_\phi \sim m^{\eta_t - \eta_\phi} \sim m^{1/\nu - 2}$ for $g=g^*$,
see Eq.~(\ref{defesponentiphi-t}), we obtain
\begin{eqnarray}
\frac{\partial^2}{\partial t^2}G^{(2)}(p,-p) &=&
G^{(4)}(0,0,p,-p) \sim m^{8-4/\nu} G^{(4)}_R(0,0,p,-p)  
\nonumber \\
&\approx& m^{8-4/\nu} \widetilde{C}(p;m) G^{(3)}_R(0,0,0) \sim (p/m)^{-1/\nu} 
  m^{d-4/\nu} \sim t^{-1-\alpha} p^{-1/\nu}.
\end{eqnarray}
Integrating this equation twice with respect to $t$, we have
\begin{equation}
\label{gam20lm}
G^{(2)}(p,-p)= a(p) + b(p) t + c t^{1-\alpha} p^{-1/\nu} + o(t^{1-\alpha}),
\end{equation}
where $a(p)$ and $b(p)$ are unknown functions of $p$. 
Comparing this result with the scaling equations 
(\ref{GE-N1}) and (\ref{Ges}), we obtain finally Eq. (\ref{lme}).
 
\subsection{The tensor correlation function}

The calculation is analogous. The short-distance expansion of the product
$T_{ij}(x) T_{ij}(y) $ is given by
\begin{equation}
T_{ij}(x+y/2)T_{ij}(x-y/2)= C_T(y) E(x) + \mbox{less singular contributions}.
\end{equation}
Now, we consider the connected correlation
function with $l$ fields $E(x)$ and two fields $T_{ij}(x)$ with the 
indices summed over,  $G_T^{(l)}(p_1,p_2;q_1,\ldots,q_l)$, 
and its renormalized counterpart
$G_{T,R}^{(l)}(p_1,p_2;q_1\ldots,q_l) = Z_T^2 Z_t^l Z_\phi^{-l-2}
G_T^{(l)}(p_1,p_2;q_1\ldots,q_l)$. 
For $p\gg m$ we have
\begin{equation}
G^{(l)}_{T,R} (p,-p;0,\ldots, 0) \approx \widetilde{C}_T(p;m) 
G^{(l+1)}_R(0,\ldots,0),
\end{equation}
The coefficient $\widetilde{C}_T(p;m)$ scales as in Eq. (\ref{scaling-Ctilde})
and satisfies the RG equation
\begin{equation}
\left[m{\partial\over \partial m} - 2 \eta_2'+ \eta_2\right] 
      \widetilde{C}_T(p;m) = 0,
\end{equation}
where $\eta_2' = \eta_T-\eta_\phi$.
Therefore
\begin{equation}
\widetilde{C}_T(p;m) \sim m^{-2} (p/m)^{-2-2\eta_2'+\eta_2} =
       m^{-2} (p/m)^{-(1+\gamma_T-\alpha)/\nu}.
\end{equation}
As in the energy case, we consider
the second derivative of $G^{(0)}_T(p,-p)$ with respect to $t$. 
For $p\gg m$ we have 
\begin{eqnarray}\label{boh}
\frac{\partial^2}{\partial t^2}G^{(0)}_T(p,-p) &= &
G^{(2)}_{T}(p,-p;0,0)\sim 
m^{8-2/\nu-2\phi_T/\nu} G^{(2)}_{T,R}(p,-p;0,0)  \nonumber \\
&\approx& 
m^{8-2/\nu-2\phi_T/\nu} \widetilde{C}(p;m) G^{(3)}_{R}(0,0,0)  
\sim p^{-(1+\gamma_T-\alpha)/\nu} t^{-1-\alpha},
\end{eqnarray}
where we have used the fact that, for $g=g^*$,
$Z_T/Z_\phi \sim m^{\eta_T - \eta_\phi} \sim m^{\phi_T/\nu - 2}$,
see Eqs.~(\ref{defesponentiphi-t}), (\ref{defesponenteetaT}).
Integrating this equation twice  with respect to $t$ and using the 
scaling equation (\ref{GT}), we obtain the large-momentum behavior 
(\ref{lmt}). 

\section{Perturbative expansion of the two-point functions  $G_{E,T}$}
\label{diag}

\begin{figure}[tb]
\hspace{-1cm}
\vspace{0cm}
\centerline{\psfig{width=12truecm,angle=0,file=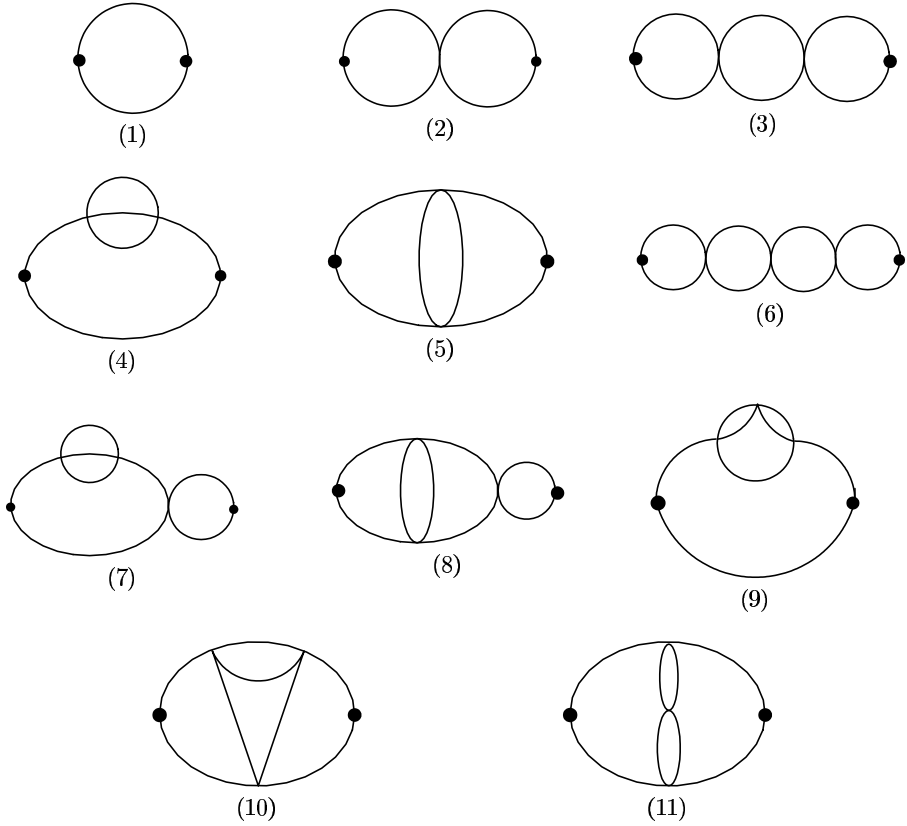}}
\vspace{0.5cm}
\caption{
Feynman diagrams contributing in the four-loop 
computation of the two-point functions $G_{E,T}$.
The black blobs indicate the insertion of the bilinear operators.
}
\label{diagrams}
\end{figure}

In order to compute the structure factor of the bilinear fields, we determine
the one-particle-irriducible diagrams 
with insertions of two operators $E$ or $T_{ij}$
and zero external legs. We use the susceptibility $\chi$ 
as inverse mass square, so that tadpole diagrams can be 
neglected. Also, subdiagrams that correspond to diagrams 
of the two-point function $\langle \phi\phi\rangle$ 
are subtracted at zero momentum.
The diagrams contributing up to four loops are
drawn in Fig.~\ref{diagrams}.
The structure factors of the bilinear fields can be expanded 
as 
\begin{equation}
{\cal G}(\bar{g},y) = u G_{E,T}(u,q)= 
\sum_{j=1} (-1)^{l-1} u^l \chi^{l/2}
S_j C_j^{E,T}  I_j(q^2 \chi)
\label{lll}
\end{equation}
where the sum is over the graphs without tadpoles,  
$l$ is the number of loops of the graph, $S_j$ the graph symmetry
factor, $C_j^{E,T}$ the group factor, and 
$I_j(q^2)$ the loop integral with unit mass.
In Table \ref{tabint} we report $S_j$, $C_j^{E,T}$.

We computed the coefficients $\bar{e}_i$ and 
$\bar{a}_i$ to four loops in the fixed-dimensions expansion.
The expansion of the loop integrals $I_j(q^2)$ is reported 
in Table \ref{tabint}. In the calculation we used the results of Refs.
\cite{Nickel-77,Rajantie-96}. 
\begin{table}[btp]
\caption{
For each diagram $j$ contributing to the energy and tensor two-point
function we report: the number of loops $l$, the symmetry factor
$S_j$, the group factors $C^{E,T}_j$, and 
the expansion of the integral $I_j(y)$ in fixed dimension $d=3$.}
\label{tabint}
\begin{tabular}{cccccl}
\multicolumn{1}{c}{$j$}&
\multicolumn{1}{c}{$l$}&
\multicolumn{1}{c}{$S_j$}&
\multicolumn{1}{c}{${C^E_j\over N}$}&
\multicolumn{1}{c}{${4 C^T_j \over N(N-1)}$}&
\multicolumn{1}{c}{$(8 \pi)^l I_j(y)$}\\
\tableline \hline
1    & 1 &   2    &   1   & 2   &$\frac{2}{\sqrt{y}} \arctan \frac{\sqrt{y}}{2}$\\
2    & 2 &   1  & ${2+N\over 3}$ & ${4\over 3}$ & $I_1^2$\\
3    & 3 &  $ {1\over 2}$  & $ {(2 + N)^2 \over 9}$ & ${8\over 9}$ &$I_1^3$\\
4    & 3 &  $\frac{2}{3}$   & $ {2+N\over 3}$& ${2(2 + N)\over 3}$&
$-0.0376821 + 0.00160802 y + 0.000209975 y^2  - 0.00012236 y^3 $ \\  
&&&&& $ \;\; + 0.0000404108 y^4  - 0.0000116689 y^5 +O(y^6)$\\
5    & 3 &  1   &  $ {2+N\over 3}$ & ${2(6+N) \over 9}$&
$0.5 - 0.105903 y + 0.0222193 y^2  - 0.00477681 y^3  $\\
&&&&& $\;\;+0.00104957 y^4  - 0.000234591 y^5+O(y^6)$ \\
6     &  4  & ${1\over 4}$   & ${(2 + N)^3 \over 27}$ & $ {16 \over 27}$ &
$I_1^4$\\
7    & 4 & $\frac{2}{3}$& ${(2 + N)^2 \over 9}$ & $ {4(2 + N)\over 9}$ &
$I_1 \; I_4$\\
8     & 4 &  1  & ${(2 + N)^2 \over 9}$ & $ {4(6 +N)\over 27}$ &
$I_1\; I_5$\\
9     & 4 &  1  & ${(2+N)(8+N)\over 27}$ & $ {2(2 + N)(8+N)\over 27}$ &
$ -0.0266277 + 0.0012789 y + 0.000157474 y^2  - 0.000095736 y^3 $\\  
&&&&& $\;\;+ 0.000031929 y^4  - 9.26022 \, 10^{-6}   y^5 +O(y^6)$\\
10    & 4 &  2  & ${(2+N)(8+N)\over 27}$& $ {8(4+N) \over 27}$ &
$ 0.25 - 0.0601852 y + 0.0132661 y^2  - 0.00292294 y^3  $\\
&&&&& $ \;\;+0.000651627 y^4  - 0.000147057 y^5 +O(y^6)$\\
11    & 4 &   ${1\over 2}$  & ${(2+N)(8+N) \over 27}$ & $ {2(16+6N + N^2)\over 27}$ &
$ 0.322467 - 0.0786798 y + 0.0175558 y^2  - 0.0039039 y^3  $\\
&&&&& $ \;\;+0.0008763 y^4  - 0.000198793 y^5  +O(y^6)$\\
\end{tabular}
\end{table}
We also used the expression of the 
bare coupling $u$ as a function of the renormalized coupling $\bar{g}$,
\begin{eqnarray}
u  =  m \,{48\pi \bar{g} \over (8+N)} \Bigl[&& 
1 + \bar{g} + \frac{  27 N^2 + 350 N + 1348 } 
    {27{\left( 8 + N \right) }^2} \,\bar{g}^2 \nonumber \\
&&+{N^3 + 17.3632 N^2 + 120.783 N + 315.831\over (8+N)^3} \bar{g}^3 + 
O(\bar{g}^4)\Bigr]
\end{eqnarray}
and the relation between $\chi$ and $m$:
\begin{equation}
m^2 \,\chi = Z_{\phi}(g) = 
1 - \frac{4 (N+2)}{27 (N+8)^2}\bar{g}^2\,-
   0.106993 \frac{N+2}{(N+8)^2} \,\bar{g}^3\, + 
O(\bar{g}^4).
\end{equation}
Writing 
\begin{eqnarray}
&&\bar{e}_i= \sum_{j=0} \bar{e}_{i,j} \bar{g}^j, \label{p1}\\
&&\bar{a}_i= \sum_{j=0} \bar{a}_{i,j} \bar{g}^j, \label{p2}
\end{eqnarray}
we computed the coefficients $\bar{e}_{i,j}$ and $\bar{a}_{i,j}$ up to $j=3$. 
They are reported in Table~\ref{eaij}.

\begin{table}[tbp]
\caption{
Coefficients $\bar{e}_{i,j}$ and $\bar{a}_{i,j}$,
cf. Eqs. (\ref{p1}) and (\ref{p2}).
}
\label{eaij}
\begin{tabular}{cccc}
\multicolumn{1}{c}{$i$}&
\multicolumn{1}{c}{$j$}&
\multicolumn{1}{c}{$\bar{e}_{i,j} (8+N)^j/(2+N)^{(1-\delta_{j0})}$}&
\multicolumn{1}{c}{$\bar{a}_{i,j} (N+8)^j$}\\
\tableline \hline
   1 & 0 & $-\case{1}{4}$ & $-\case{1}{4}$ \\
     & 1 &$ \case{1}{6} $ &$\case{1}{3}$ \\
     & 2 & $ 0.0290678 $ &$ 0.0581358 - 0.0195433
   \, N $ \\
     & 3 &$ 0.2941 - 0.0105012 N  $ 
   & $0.588201 - 0.158869 \,N - 0.044705 \,N^2$ \\
   &\\
   2 & 0 & $\case{1}{16}$ &$\case{1}{16}$ \\
     & 1 & $-\case{1}{15} $& $ -\case{2}{15}$ \\
     & 2& $0.0280008 + 0.0208333 \,N$&$ 0.0560016 +
   0.00585959 \,N $\\
    & 3&$ -0.046127 + 0.0223387 \,N $
   & $-0.0922537 + 0.0828459 \,N + 0.0173899 \,N^2$\\
   &\\
   3& 0&$-\case{1}{64}$&$ -\case{1}{64}$ \\
    & 1&$ \case{71}{3360}$& $ \case{71}{1680} $\\
    & 2& $ -0.0179798 - 0.0114583 N $ &$ -0.0359598 -
   0.00137167 N $\\
    & 3&$ 0.00517395- 0.00217669 \,N + 0.00231481 N^2 $
   &$ 0.0103479 - 0.0333737 \, N - 0.0056229 \, N^2$ \\
   &\\
   4 & 0&$\case{1}{256}$ &$\case{1}{256}$ \\
    & 1&$-\case{31}{5040}$&$ -\case{31}{2520}$\\
    & 2&$ 0.00733893 + 0.0044494 N $ &$0.0146779 +
   0.000299779 N$ \\
    & 3&$-0.000842908 - 0.00212265 N + 0.00162037 N^2 $
   &$-0.00168582 + 0.0117321 N + 0.00170808 N^2 $\\
   &\\
   5 & 0& $-\case{1}{1024}$&$-\case{1}{1024} $\\
    & 1& $\case{3043}{1774080}$&$ \case{3043}{887040} $\\
    & 2& $-0.00253604 - 0.00149678 N $&$ -0.00507208 -
   0.0000641563 N$\\
    & 3&$-0.000390898 - 0.0015352 N - 0.000747354 N^2 $
   &$0.000781794 - 0.0037964 N - 0.000501577 N^2 $ 
\end{tabular}
\end{table}

We computed the coefficients $e_i$ and $a_i$ in $\epsilon$
expansion to three loops, i.e. to order $\epsilon^3$. The 
expansion of the integrals $I_4(y)$ and $I_5(y)$ was
obtained by using the algebraic algorithm of Ref.~\cite{PV-00}. 

\begin{table}
\caption{Expansion coefficients $x_1$, $x_2$ and $x_3$ for $e_i$ and $a_i$.}
\label{eps-smallmomentum-1} 
\begin{center}
\begin{tabular}{cccc}
& $x_1$ & $x_2$ & $x_3$ \\
\tableline \hline
${e}_1$  & 
${\frac{1}{12}}$  & 
${\frac{\left( 2 + N \right) \,\left( 44 + 13\,N \right) }{12}}$  & 
$-{\frac{7}{72}}$ \\
${e}_2$  & 
$-{\frac{1}{120}}$  &
${\frac{-40 - 270\,N - 42\,{N^2} + {N^3}}{360}}$  &
${\frac{7}{360}}$ \\
${e}_3$  & 
${\frac{1}{840}}$  &
${\frac{-352 + 1176\,N + 180\,{N^2} - 5\,{N^3}}{10080}}$  &
$-{\frac{1}{240}}$ \\
${e}_4$  & 
$-{\frac{1}{5040}}$  &
${\frac{1808 - 3072\,N - 477\,{N^2} + 13\,{N^3}}{151200}}$  &
${\frac{1}{1080}}$ \\
${e}_5$  & 
${\frac{1}{27720}}$  &
${\frac{-3392 + 4208\,N + 668\,{N^2} - 17\,{N^3}}{1108800}}$  &
$-{\frac{1}{4752}}$ \\
\hline
${a}_1$  & 
$-{\frac{1}{12}}$  &
${\frac{22 - N}{12}}$  &
${\frac{7}{72}}$ \\
${a}_2$  & 
${\frac{1}{120}}$  &
${\frac{-20 + 44\,N + 3\,{N^2}}{720}}$  &
$-{\frac{7}{360}}$ \\
${a}_3$  & 
$-{\frac{1}{840}}$  &
${\frac{-88 - 128\,N - 9\,{N^2}}{10080}}$  &
${\frac{1}{240}}$ \\
${a}_4$  & 
${\frac{1}{5040}}$  &
${\frac{904 + 778\,N + 55\,{N^2}}{302400}}$  &
$-{\frac{1}{1080}}$ \\
${a}_5$  & 
$-{\frac{1}{27720}}$  &
${\frac{-2544 - 1768\,N - 125\,{N^2}}{3326400}}$  &
${\frac{1}{4752}}$ \\
\end{tabular}
\end{center}
\end{table}

\begin{table}
\caption{Expansion coefficients $x_4$ and $x_5$ for $e_i$ and $a_i$.}
\label{eps-smallmomentum-2} 
\begin{tabular}{ccc}
${e}_1$  & $x_4$ &
${\frac{-1184 - 348\,N - 7\,{N^2}}{32}}$  \\
& $x_5$ & ${\frac{-\left( 2 + N \right) \,
        \left( -4112 - 2596\,N - 466\,{N^2} + {N^3} \right)}{24}}$ \\
${e}_2$  & $x_4$ &
${\frac{4512 + 1420\,N + 35\,{N^2}}{1280}}$  \\
& $x_5$ & ${\frac{\left( 2 + N \right) \,\left( -170624 - 78048\,N - 8148\,{N^2} + 
        509\,{N^3} \right) }{5760}} $ \\
${e}_3$  & $x_4$ &
${\frac{-2311264 - 726900\,N - 17885\,{N^2}}{4587520}}$  \\ 
& $x_5$ & ${\frac{586167808 + 604936960\,N + 160060792\,{N^2} + 8862854\,{N^3} - 
      1077019\,{N^4} + 2048\,{N^5}}{61931520}}$ \\
${e}_4$  & $x_4$ &
${\frac{2370848 + 719452\,N + 16023\,{N^2}}{27525120}}$  \\
& $x_5$ & ${\frac{-1006341632 - 1133004544\,N - 290766168\,{N^2} - 13609246\,{N^3} + 
      2081479\,{N^4} - 3072\,{N^5}}{619315200}}$ \\
${e}_5$  & $x_4$ &
${\frac{-2525643296 - 724481980\,N - 13348335\,{N^2}}{155021475840}}$ \\
& $x_5$ & ${\frac{3084090980864 + 3778414828800\,N + 961368439624\,{N^2} + 
      41380542442\,{N^3} - 6987518685\,{N^4} + 2097152\,{N^5}}{10463949619200}}
$\\
\hline
${a}_1$  & $x_4$ &
${\frac{-1376 + 764\,N + 301\,{N^2} + 14\,{N^3}}{64}}$ 
\\
& $x_5$ & ${\frac{7552 - 2784\,N - 5692\,{N^2} - 1667\,{N^3} - 100\,{N^4}}{96}}$ \\
${a}_2$  & $x_4$ &
${\frac{-\left( 3104 + 12300\,N + 3185\,{N^2} + 140\,{N^3} \right) }
    {5120}}$  \\
& $x_5$ & ${\frac{-1087744 - 484736\,N + 26236\,{N^2} + 32387\,{N^3} + 2316\,{N^4}}
    {23040}}$ \\
${a}_3$  & $x_4$ &
${\frac{1445632 + 1966400\,N + 440020\,{N^2} + 17885\,{N^3}}
    {4587520}}$  \\
& $x_5$ & ${\frac{568281088 + 262815488\,N - 1452256\,{N^2} - 13472924\,{N^3} - 
      1114341\,{N^4} - 9216\,{N^5}}{61931520}}$ \\
${a}_4$  & $x_4$ &
${\frac{-\left( 33881888 + 34540972\,N + 7018809\,{N^2} + 256368\,{N^3}\
        \right) }{440401920}}$ \\ 
& $x_5$ & ${\frac{-17032726784 - 7719736448\,N + 90484812\,{N^2} + 409316471\,{N^3} + 
      36197264\,{N^4} + 491520\,{N^5}}{9909043200}}$ \\
${a}_5$  & $x_4$ &
${\frac{5244438496 + 4565788340\,N + 853347495\,{N^2} + 26696670\,{N^3}}
    {310042951680}}$  \\
& $x_5$ & ${\frac{34179580035328 + 14923763131008\,N - 513765484268\,{N^2} - 
      895119755607\,{N^3} - 81461592926\,{N^4} - 
      1376256000\,{N^5}}{104639496192000}}$ \\
\end{tabular}
\end{table}

We write
\begin{eqnarray}
e_i &=& {N-4\over N+8} x_1 \epsilon + 
      {1\over (N+8)^3} x_2 \epsilon^2 + 
\nonumber \\ 
&&
      \left[
      {(N+2)(N-4)\over (N+8)^3} x_3 \lambda + 
      {(N+2)\over (N+8)^4} x_4 \zeta(3) + 
      {1\over (N+8)^5} x_5 \right]\epsilon^3 + O(\epsilon^4), \\
a_i &=& {N+4\over N+8} x_1 \epsilon + 
      {N+4\over (N+8)^3} x_2 \epsilon^2 + 
\nonumber \\
&&
      \left[
      {(N+2)(N+4)\over (N+8)^3} x_3 \lambda + 
      {1\over (N+8)^4} x_4 \zeta(3) + 
      {1\over (N+8)^5} x_5 \right]\epsilon^3 + O(\epsilon^4), 
\end{eqnarray}
where $\lambda = 1.171953619344729445$. 
The coefficients $x_i$ are reported in Tables
\ref{eps-smallmomentum-1} and \ref{eps-smallmomentum-2}.
Note that \cite{NA-97}  to order $\epsilon^2$, 
$e_1 \approx -\alpha/(6\gamma)$ and 
$a_1 \approx -\gamma_T/(6\gamma)$. These relations do not hold at 
order $\epsilon^3$.


\begin{thebibliography}{199}
\bibitem{foot1}
In the analyses we used the following
estimates of $\bar{g}^*$:
$\bar{g}^*=1.405(2)$ for $N=1$ (from Ref. \cite{review}),
$\bar{g}^*=1.402(4)$ for $N=2$ (from Ref. \cite{CHPRV-01}),
$\bar{g}^*=1.395(7)$ for $N=3$ (from Ref. \cite{CHPRV-02}),
$\bar{g}^*=1.385(15)$, $1.360(15)$,
$1.310(15)$, $1.205(15)$ for $N=4$, 5, 8, 16 respectively
(from Refs. \cite{GZ-98,BC-98,PV-00}).
Other numerical estimates can be found in Ref. \cite{review}.
\bibitem{foot2}
For the exponent $\nu$ we have taken the following numerical
results:
$\nu=0.63002(23)$ for $N=1$ (Ref. \cite{CPRV-99}),
$\nu=0.67155(27)$ for $N=2$ (Ref. \cite{CHPRV-01}),
$\nu = 0.7112(5)$ for $N=3$ (Ref. \cite{CHPRV-02}),
$\nu = 0.749(2)$  for $N=4$ (Ref. \cite{Hasenbusch-01}),
$\nu = 0.766$,
$0.830$, $0.911$ for $N=5,8,16$ (Ref. \cite{AS-95}).
Other numerical estimates can be found in Ref. \cite{review}.
\bibitem{T23}
The same singular behavior for $N\rightarrow 4$
is present in the
$\epsilon$-expansion of the coefficients 
$A_i$ of the 
large-momentum expansion of the fundamental 
two-point function, cf. Eq.~(\ref{FL-law}).
Indeed, one finds \cite{FA-73}
$A_1  = 1 + O(\epsilon)$, 
$A_2=\case{2+N}{4-N} + O(\epsilon)$, and
$A_3 = - \case{6}{4-N} + O(\epsilon)$.
\end{thebibliography}
\end{document}